\documentclass[12pt, draftclsnofoot, onecolumn]{IEEEtran}

\usepackage{amsmath}
\usepackage{amssymb}
\usepackage{amsfonts}
\usepackage{graphicx}
\usepackage{epsfig}
\usepackage{subfigure}
\usepackage{psfrag}
\usepackage{cite}
\usepackage{latexsym}
\usepackage{url}
\usepackage{color}
\usepackage{multirow}

\PassOptionsToPackage{bookmarks={false}}{hyperref}

\begin{document}
\title{Transmit Optimization for Symbol-Level Spoofing
}
\author{Jie Xu, Lingjie Duan, and Rui Zhang\\
\thanks{Part of this paper has been submitted to the IEEE Global Communications Conference (GLOBECOM) workshop on Trusted Communications with Physical Layer Security (TCLPS), Washington, DC USA, December 4-8, 2016 \cite{GCworkshop}.}
\thanks{J. Xu is with the School of Information Engineering, Guangdong University of Technology (e-mail: jiexu.ustc@gmail.com). He is also with the Engineering Systems and Design Pillar, Singapore University of Technology and Design.}
\thanks{L. Duan is with the Engineering Systems and Design Pillar, Singapore University of Technology and Design (e-mail:~lingjie\_duan@sutd.edu.sg).}
\thanks{R. Zhang is with the Department of Electrical and Computer Engineering, National University of Singapore (e-mail: elezhang@nus.edu.sg). He is also with the Institute for Infocomm Research, A*STAR, Singapore.}
}

\maketitle

\begin{abstract}
With recent developments of wireless communication technologies, malicious users can use them to commit crimes or launch terror attacks, thus imposing new threats on the public security. To quickly respond to defend these attacks, authorized parities (e.g., the National Security Agency of the USA) need to intervene in the malicious communication links over the air. This paper investigates this emerging wireless communication intervention problem at the physical layer. Unlike prior studies using jamming to disrupt or disable the targeted wireless communications, we propose a new physical-layer spoofing approach to change their communicated information. Consider a fundamental three-node system over additive white Gaussian noise (AWGN) channels, in which an intermediary legitimate spoofer aims to spoof a malicious communication link from Alice to Bob, such that the received message at Bob is changed from Alice's originally sent message to the one desired by the spoofer. We propose a new symbol-level spoofing scheme, where the spoofer designs the spoofing signal via exploiting the symbol-level relationship between each original constellation point of Alice and the desirable one of the spoofer. In particular, the spoofer aims to minimize the average spoofing-symbol-error-rate (SSER), which is defined as the average probability that the symbols decoded by Bob fail to be changed or spoofed, by designing its spoofing signals over symbols subject to the average transmit power constraint. By considering two cases when Alice employs the widely-used binary phase-shift keying (BPSK) and quadrature phase-shift keying (QPSK) modulations, we obtain the respective optimal solutions to the two average SSER minimization problems. Numerical results show that the symbol-level spoofing scheme with optimized transmission achieves a much lower average SSER, as compared to other benchmark schemes.
\end{abstract}

\newtheorem{definition}{\underline{Definition}}[section]
\newtheorem{fact}{Fact}
\newtheorem{assumption}{Assumption}
\newtheorem{theorem}{\underline{Theorem}}[section]
\newtheorem{lemma}{\underline{Lemma}}[section]
\newtheorem{corollary}{\underline{Corollary}}[section]
\newtheorem{proposition}{\underline{Proposition}}[section]
\newtheorem{example}{\underline{Example}}[section]
\newtheorem{remark}{\underline{Remark}}[section]
\newtheorem{algorithm}{\underline{Algorithm}}[section]
\newcommand{\mv}[1]{\mbox{\boldmath{$ #1 $}}}

\section{Introduction}

Recent technological advancements have enabled increasing use of infrastructure-free wireless communications. For example, smartphone users can exchange information with each other by exploiting local Wi-Fi and Bluetooth connections, or using the fifth-generation (5G) cellular device-to-device communications; and even unmanned aerial vehicles (UAVs) can directly communicate with nearby ground stations and send back photos and videos in real time. Although these infrastructure-free communication links bring great convenience to our daily lives, they can also be used by malicious users to launch various security attacks. For instance, terrorists can use peer-to-peer Wi-Fi connections to communicate and facilitate terror attacks, and criminals can control UAVs to spy and collect private information from rightful users. As such malicious attacks are launched via infrastructure-free wireless communications, they are difficult to be monitored by solely using existing information surveillance methods that intercept the communication data at the cellular or Internet infrastructures.{\footnote{See, e.g., the Terrorist Surveillance Program launched by the National Security Agency in the USA at \url{https://nsa.gov1.info/surveillance/}.}} In response to such new threats on public security, authorized parties such as government agencies should develop new approaches to legitimately surveil these suspicious wireless communication links over the air (e.g., via eavesdropping) to detect malicious attacks, and then intervene in them (e.g., via jamming and spoofing) to quickly defend and disable these attacks.

There have been several recent studies in the literature that investigate the surveillance of wireless communications, where authorized parties efficiently intercept suspicious wireless communication links, extract their exchanged data contents, and help identify the malicious wireless communication links to intervene in. Conventionally, the methods for wireless communications surveillance include wiretapping of wireless operators' infrastructures and installation of monitoring software in smartphones. Recently, over-the-air eavesdropping has emerged as a new wireless communications surveillance method. Among others, passive eavesdropping (see, e.g., \cite{ZouWangHanzo2015}) and proactive eavesdropping \cite{XuDuanZhang1,XuDuanZhang2,ZengZhang,ZengZhang2} are two approaches implemented at the physical layer, in which authorized parties can deploy dedicated wireless monitors to overhear the targeted wireless communications, especially the infrastructure-free ones.

Efficient surveillance can help detect and identify malicious users and their communications. After that, authorized parties need to quickly respond and defend them via wireless communication intervention. For example, the security agency may need to disrupt, disable, or spoof ongoing terrorists' communications to prevent terror attacks at the planning stage, and it is also desirable to change the control signal of a malicious UAV to land it in a targeted location and catch it. In the literature, physical-layer jamming (see, e.g., \cite{Liu2015,Bayesteh2004,Brady2006,Rodrigues2009,Jorswieck2005,Medard,Kashyap2004,Shafiee2009}) is one existing approach that can be employed to intervene in malicious communications, though it was originally proposed for military instead of public security applications. In the physical-layer jamming, the jammer sends artificially generated Gaussian noise (so-called ``uncorrelated jamming'' \cite{Liu2015,Bayesteh2004,Brady2006,Rodrigues2009,Jorswieck2005}) or a processed version of the malicious signal (so-called ``correlated jamming'' \cite{Medard,Kashyap2004,Shafiee2009}) to disrupt or disable the targeted malicious wireless communications. However, jamming the targeted communications at the physical layer is easy to be detected, and may not be sufficient to successfully intervene in malicious activities. This is due to the fact that when the targeted communication continuously fails due to the jamming attack, the malicious users may take counter-measures by changing their communication frequency bands or switching to another way of communications. Thus, we are motivated to study a new wireless communication intervention via spoofing at the physical layer, which can keep the malicious communication but change the communicated information to intervene in.

\begin{figure}
\centering
 \epsfxsize=1\linewidth
    \includegraphics[width=8cm]{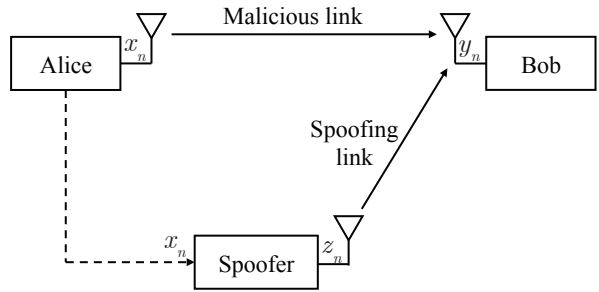}
\caption{The system model with a spoofer aiming to purposely change the information content transmitted from Alice to Bob.} \label{fig:1}\vspace{-2em}
\end{figure}

\begin{figure}
\centering
 \epsfxsize=1\linewidth
    \includegraphics[width=8cm]{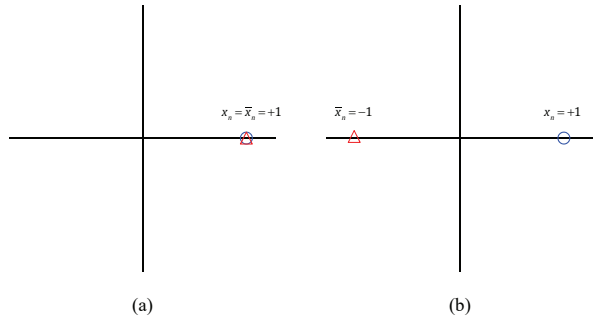}
\caption{Illustration of different types of symbols with the BPSK modulation, where the red triangle denotes the original constellation point $x_n$ of Alice, and the blue circular denotes the desirable constellation point $\bar x_n$ of the spoofer. (a) An example of Type-I symbols, where  $x_n$ and $\bar x_n$ are identical with $x_n = \bar x_n = +1$; (b) An example of Type-II symbols, where $x_n$ and $\bar x_n$ are opposite with $x_n = +1$ and $\bar x_n = -1$.} \label{fig:BPSK}\vspace{-0em}
\end{figure}

We investigate the new physical-layer spoofing by considering a fundamental three-node system over additive white Gaussian noise (AWGN) channels. As shown in Fig. \ref{fig:1},  an intermediary legitimate spoofer aims to spoof a malicious communication link from Alice to Bob, such that the received message at Bob is changed from Alice's originally sent message to the one desired by the spoofer. Under this setup, we propose a new symbol-level spoofing approach, in which the spoofer designs the spoofing signals via exploiting the symbol-level relationship between each original constellation point of Alice and the desirable one of the spoofer, so as to optimize the spoofing performance. In particular, we consider two cases when Alice employs the widely-used binary phase-shift keying (BPSK) and quadrature phase-shift keying (QPSK) modulations, respectively.{\footnote{Note that the symbol-level spoofing approach is extendible to other modulation techniques such as $M$-ary quadrature amplitude modulation ($M$-QAM) and $M$-ary phase shift keying ($M$-PSK) with $M>4$. Nevertheless, under these modulation techniques, how to design spoofing signals to optimally solve the average SSER minimization problem is generally a more difficult task, since the corresponding SSER functions will become very complicated.}} The objective of the spoofer is to minimize the average spoofing-symbol-error-rate (SSER), i.e., the average probability that the symbols decoded by Bob fail to be changed as the desirable ones of the spoofer. The main results of this paper are summarized as follows.
\begin{itemize}
\item In the BPSK case (with the constellation points being $\pm 1$), the spoofing signals are designed by classifying the symbols into two types. In each of Type-I symbols (see Fig. \ref{fig:BPSK}-(a)), where the original constellation point of Alice and the desirable one of the spoofer are identical (both are $+1$ or $-1$), the spoofing signal is designed to {\it constructively} combine with the original signal of Alice at Bob to help improve the decoding reliability against Gaussian noise. In each of Type-II symbols (see Fig. \ref{fig:BPSK}-(b)), where the original constellation point of Alice and the desirable one of the spoofer are opposite (one is $+1$ (or $-1$) but the other is $-1$ (or $+1$)), the spoofing signal is designed to {\it destructively} combine with the original signal of Alice at Bob, thus moving the constellation point towards the desirable opposite direction. We minimize the average SSER by optimizing the spoofing signals and their power allocations over Type-I and Type-II symbols at the spoofer, subject to its average transmit power constraint. Although this problem is non-convex, we derive its optimal solution. It is shown that when the transmit power at Alice is low or the spoofing power at the spoofer is high, the spoofer should allocate its transmit power to both Type-I and Type-II symbols. Otherwise, when the transmit power at Alice is high and the spoofing power at the spoofer is low, the spoofer should allocate almost all its transmit power over a certain percentage of Type-II symbols with an ``on-off'' power control.
\item In the QPSK case with the constellation points being $(\pm 1 \pm j)/\sqrt{2}$ with $j=\sqrt{-1}$, the symbols are further classified into three types, where in Type-I, Type-II, and Type-III symbols, the original constellation points of Alice and the desirable ones of the spoofer are identical, opposite, and neighboring, respectively, as shown in Fig. \ref{fig:QPSK}. For Type-I and Type-II symbols, the spoofing signals are designed to have equal strengths for the real and imaginary components, such that at the receiver of Bob they can be be constructively and destructively combined with the original constellation points by Alice, respectively. For Type-III symbols, the spoofing signals are designed to have independent real and imaginary components. Under such a design, we formulate the average SSER minimization problem by optimizing the spoofing power allocations over symbols, subject to the average transmit power constraint. Though this problem is non-convex and generally difficult, we obtain its optimal solution, motivated by that in the BPSK case.
\item Numerical results show that for both BPSK and QPSK cases, the symbol-level spoofing scheme with optimized transmission achieves a much better spoofing performance (in terms of a lower average SSER), as compared to the block-level spoofing benchmark where the spoofer does not exploit the symbol information of Alice, and a heuristically designed symbol-level spoofing scheme.
\end{itemize}

It is worth noting that in the existing literature there is another type of higher-layer spoofing attack, which can also be utilized for wireless communication intervention (see, e.g., \cite{ZouWangHanzo2015,Nagarajan2010,CERT1995,Kannhavong2007}). For example, in the medium access control (MAC) spoofing \cite{Nagarajan2010} and Internet protocol (IP) spoofing, a network attacker can hide its true identity and impersonate another user, so as to access the targeted wireless networks. Nevertheless, for these higher-layer spoofing, the network attacker needs to establish new wireless communication links to access the network. In contrast, our proposed symbol-level spoofing is implemented at the physical layer, which can change the communicated information of {\it ongoing} malicious wireless communications, thus leading to a quicker response and intervention that is also more likely to be covert.

It is also worth comparing our proposed symbol-level spoofing versus the symbol-level precoding (not for security) in downlink multiuser multi-antenna systems \cite{Masouros2009,Alodeh2015}. In the symbol-level precoding, the transmitter designs its precoding vectors by exploiting the symbol-level relationships among the messages to different receivers, such that the constructive part of the inter-channel interference is preserved and exploited and only the destructive part is eliminated. Although the symbol-level spoofing and precoding are based on a similar design principle of exploiting the symbol-level relationship among co-channel signals, they focus on different application scenarios for different purposes, thus requiring different design methods.

The remainder of this paper is organized as follows. Section \ref{sec:2} introduces the system model and formulates the average SSER minimization problem. Sections \ref{sec:3} and \ref{sec:QPSK} propose the symbol-level spoofing approach and design the spoofing signals and their power allocations for the cases of BPSK and QPSK modulations, respectively. Section \ref{sec:4} presents numerical results to evaluate the performance of the proposed symbol-level spoofing design as compared to other benchmark schemes. Finally, Section \ref{sec:5} concludes the paper.

\section{System Model and Problem Formulation}\label{sec:2}

As shown in Fig. \ref{fig:1}, we consider a fundamental three-node system over AWGN channels, where  an intermediary  legitimate spoofer aims to spoof a malicious wireless communication link from Alice to Bob by changing the communicated data at the Bob side. We consider that the malicious communication employs the BPSK or QPSK modulation techniques, which are most commonly used in existing wireless communication systems. In the $n$th symbol of this block, we denote the transmitted signal by Alice as $\sqrt{P}x_n$, where $P$ is the transmit power per symbol at Alice, and $x_n$ denotes the message that Alice wants to deliver to Bob. Here, $x_n$ is equally likely chosen from the set of constellation points $\mathcal M$, where $\mathcal{M} = \{\pm1\}$ and $\mathcal{M} = \{(\pm1  \pm j)/{\sqrt{2}}\}$ for the BPSK and QPSK cases, respectively. Therefore, we have $|x_n|^2 = 1$.

First, we introduce the receiver model of Bob by considering the case without spoofing. Accordingly, the received signal by Bob in the $n$th symbol is expressed as
\begin{align}
r_n = \sqrt{P}x_n+v_n,
\end{align}
where $v_n$ denotes the noise at the receiver of Bob, which is an independent and identically distributed (i.i.d.) circularly symmetric complex Gaussian (CSCG) random variable with zero mean and unit variance. Based on the maximum likelihood (ML) detection, the decoded message by Bob is expressed as
\begin{align}
\arg \min_{s \in \mathcal{M}} |r_n - \sqrt{P} s|^2.
\end{align}

Next, we consider the spoofing strategy employed by the spoofer. It is assumed that the spoofer perfectly knows the transmitted symbol information $x_n$'s of Alice. Here, $x_n$'s can be practically obtained by the spoofer via efficient eavesdropping or wiretapping beforehand. For example, if Alice is an intermediary node of a multi-hop communication link, then the spoofer can obtain $x_n$'s via eavesdropping the previous hops; if Alice gets its transmitted data from the backhaul or infrastructure-based networks, then the spoofer can acquire them via using wiretapping devices to overhear the backhaul communications; and furthermore, the spoofer can even secretly install an interceptor software (e.g., FlexiSPY\footnote{See {\url{http://www.flexispy.com/}}.}) in the Alice's device to get $x_n$'s. Note that the assumption about the perfect symbol information at the spoofer has been made in the existing correlated jamming literature (see, e.g.,\cite{Medard,Kashyap2004}) to improve the jamming performance. We make a similar assumption here for the purpose of characterizing the spoofing performance upper bound, and leave the details about the symbol information acquisition for future work. Based on the information of $x_n$'s, the spoofer designs the spoofing signal as $z_n$ in the $n$th symbol (the design details will be provided in the next section). Then, the received signal at Bob is expressed as
\begin{align}
y_n =  \sqrt{P}x_n+z_n+v_n.\label{eqn:y_n}
\end{align}
With the ML detection, the decoded message by Bob is expressed as
\begin{align}
\hat x_n = \arg \min_{s \in \mathcal{M}} |y_n -  \sqrt{P}s|^2.\label{eqn:hatx_n}
\end{align}


The spoofer aims to maximize the opportunity of changing the messages of Alice to be the desirable ones by itself. Let $\bar x_n$ denote the desirable constellation point for the $n$th symbol, which is equally likely chosen from $\mathcal M$ and is independent from the message $x_n$ sent by Alice. Nevertheless, due to the limited spoofing power and receiver noise, it is difficult for the spoofer to ensure that all symbols $\hat x_n$'s are successfully changed to be the desirable $\bar x_n$'s. In this case, we define the probability of unsuccessful spoofing in any symbol $n$ as the SSER, denoted by $\mathrm{Pr}(\hat x_n \neq \bar x_n)$.{\footnote{Note that with BPSK, the SSER is equivalent to the spoofing-bit-error-rate (SBER).}} Then, the objective of the spoofer is to minimize the average SSER, i.e., $\mathbb{E}_n\left(\mathrm{Pr}(\hat x_n \neq \bar x_n)\right)$, where $\mathbb{E}_n(\cdot)$ denotes the statistical expectation over all possible symbols. Suppose that the spoofer is constrained by a maximum average transmit power denoted by $Q$, i.e., $\mathbb{E}_n(|z_n|^2) \le Q$. As a result, the optimization problem of our interest is
\begin{align}
\min_{\{z_n\}}~&\mathbb{E}_n\left(\mathrm{Pr}(\hat x_n \neq \bar x_n)\right)\nonumber\\
\mathrm{s.t.}~&\mathbb{E}_n(|z_n|^2) \le Q.\label{eqn:5}
\end{align}

In the following two sections, we will solve problem (\ref{eqn:5}) by considering the BPSP and QPSK modulations, respectively.

\section{Optimal Symbol-Level Spoofing Design with BPSK Signaling}\label{sec:3}

In this section, we consider the case with BPSK signaling, i.e., $\mathcal{M} = \{\pm1\}$. In the following, we first propose the symbol-level spoofing signals design and then optimally solve the average SSER minimization problem (\ref{eqn:5}) in this case.

\subsection{Spoofing Signals Design and Problem Reformulation}
To facilitate the description, as shown in the examples in Fig. \ref{fig:BPSK}, we classify the symbols over each block into two types as follows based on the relationship between the original constellation point $x_n$ of Alice and the desirable one $\bar x_n$ of the spoofer in each symbol $n$.
\begin{itemize}
  \item {\it Type-I symbol}: The symbol $n$ is called a Type-I symbol if $x_n$ and $\bar x_n$ are identical ($x_n = \bar x_n = +1$ or $x_n = \bar x_n = -1$). We denote the set of all Type-I symbols as $\mathcal N_1$.
  \item {\it Type-II symbol}: The symbol $n$ is called a Type-II symbol if $x_n$ and $\bar x_n$ are opposite ($x_n = +1$ and $\bar x_n = -1$, or $x_n = -1$ and $\bar x_n = +1$). We denote the set of all Type-II symbols as $\mathcal N_2$.
\end{itemize}
In the following two propositions, we present the optimal symbol-level spoofing signal design, and obtain the corresponding SSER functions.
\begin{proposition}\label{proposition:TypeI}
Given any Type-I symbol $n\in\mathcal N_1$, it is optimal to minimize the conditional SSER $\mathrm{Pr}(\hat x_n \neq \bar x_n|x_n = \bar x_n)$ by designing $z_n = \sqrt{A_n}x_n $ aligning with $x_n$, where $A_n$ denotes the spoofing power for this symbol. Accordingly, $\mathrm{Pr}(\hat x_n \neq \bar x_n|x_n = \bar x_n)$ is given as
\begin{align}
f_1(A_n) = &\frac{1}{2} - \frac{1}{2}\mathrm{erf}\left(\sqrt{A_n}+\sqrt{P}\right),\label{eqn:symbol:I}
\end{align}
where $\mathrm{erf}(\cdot)$ is the error function defined as
\begin{align*}
\mathrm{erf}(x) = \frac{2}{\sqrt{\pi}}\int_{0}^{x}e^{-t^2}\mathrm{d}t.
\end{align*}
\end{proposition}
\begin{IEEEproof}
See Appendix \ref{proof:TypeI}.
\end{IEEEproof}

\begin{proposition}\label{proposition:TypeII}
Given any Type-II symbol $n\in\mathcal N_2$, it is optimal to minimize the conditional SSER $\mathrm{Pr}(\hat x_n \neq \bar x_n|x_n \neq \bar x_n)$ by designing $z_n = - \sqrt{B_n}x_n$ opposite to $x_n$, where $B_n$ denotes the spoofing power for this symbol. Accordingly, $\mathrm{Pr}(\hat x_n \neq \bar x_n|x_n \neq \bar x_n)$ is given as
\begin{align}
f_2(B_n) = &\frac{1}{2} - \frac{1}{2}\mathrm{erf}\left(\sqrt{B_n}-\sqrt{P}\right).\label{eqn:symbol:II}
\end{align}
\end{proposition}
\begin{IEEEproof}
This proposition can be proved by following a similar procedure as for Proposition \ref{proposition:TypeI}. Therefore, the details are omitted for brevity.
\end{IEEEproof}

Propositions \ref{proposition:TypeI} and \ref{proposition:TypeII} are intuitive. In each Type-I symbol, Proposition \ref{proposition:TypeI} shows that the spoofing signal should be designed such that at the receiver of Bob it is {\it constructively} combined with the original signal from Alice, thus increasing the received power of the desirable constellation point against Gaussian noise. In each Type-II symbol, Proposition \ref{proposition:TypeII} shows that at the receiver of Bob the spoofing signal should be {\it destructively} combined with the original signal from Alice, so as to move the constellation point towards the desirable opposite direction.

Based on these two propositions, the average SSER minimization problem (\ref{eqn:5}) is specified as follows by jointly optimizing the spoofing power $A_n$'s over Type-I symbols and $B_n$'s over Type-II symbols.
\begin{align}
\min_{\{A_n \ge 0\},\{B_n\ge 0\}} & \frac{1}{2}\left(\mathbb{E}_{n\in\mathcal N_1} \left(f_1(A_n)\right) + \mathbb{E}_{n\in\mathcal N_2} \left(f_2(B_n)\right)\right)\nonumber\\
\mathrm{s.t.}~~~~&~\frac{1}{2}\left(\mathbb{E}_{n\in\mathcal N_1} (A_n) + \mathbb{E}_{n\in\mathcal N_2} (B_n) \right)\le Q, \label{Problem:BPSK}
\end{align}
where the term $1/2$ follows from the fact that each of the two symbol sets $\mathcal N_1$ and $\mathcal N_2$ on average occupies a half of all symbols over each block.

The spoofing power allocation problem (\ref{Problem:BPSK}) is generally non-convex, since the SSER function $f_2(B_n)$ in the objective is non-convex over $B_n \ge 0$ (as will be shown next). Therefore, this problem is difficult to solve. In the following, we first show some useful properties of the SSER functions $f_1(A_n)$ and $f_2(B_n)$, and then present the optimal solution to problem (\ref{Problem:BPSK}).

\subsection{Properties of the SSER Functions $f_1(A_n)$ and $f_2(B_n)$}

First, we have the following lemma for the SSER function $f_1(A_n)$.
\begin{lemma}\label{proposition:1}
$f_1(A_n)$ is monotonically decreasing and convex over $A_n \in [0,+\infty)$.
\end{lemma}
\begin{IEEEproof}
It is easy to show that over $A_n \in [0,+\infty)$, the first- and second-order derivatives of $f_1(A_n)$ satisfy that $f_1'(A_n) \le 0$ and $f_1''(A_n)\ge 0$, respectively. Therefore, this lemma follows.
\end{IEEEproof}

Next, we study the SSER function $f_2(B_n)$.

\begin{lemma}\label{proposition:2}
$f_2(B_n)$ is monotonically decreasing over $B_n \in [0,+\infty)$. The convexity of $f_2(B_n)$ is given as follows depending on Alice's transmit power $P$.
\begin{itemize}
  \item {\it Alice's low transmit power regime (i.e., $P \le 2$)}: $f_2(B_n)$ is convex over $B_n \in [0,+\infty)$.
  \item {\it Alice's high transmit power regime (i.e., $P > 2$)}: $f_2(B_n)$ is first convex over $B_n \in [0,\zeta_1]$, then concave over $B_n \in(\zeta_1,\zeta_2)$, and finally convex over $B_n\in[\zeta_2,+\infty)$, where the two boundary points $\zeta_1 < \zeta_2$ are given as
\begin{align}
\zeta_1 &= \left(\frac{\sqrt{P} - \sqrt{P-2}}{2}\right)^2,\label{eqn:zeta1}\\
\zeta_2 &= \left(\frac{\sqrt{P} + \sqrt{P-2}}{2}\right)^2.\label{eqn:zeta2}
\end{align}
\end{itemize}
\end{lemma}
\begin{IEEEproof}
See Appendix \ref{proof:2}.
\end{IEEEproof}

In the Alice's high transmit power regime when $P > 2$, we further have the following property for $f_2(B_n)$.

\begin{lemma}\label{proposition:3}
When $P > 2$, there exist two points $\tau_1$ and $\tau_2$ with $0 < \tau_1 \le \zeta_1$ and $\tau_2 \ge \zeta_2$, such that all the points $(B_n,f_2(B_n))$ are above the straight line passing through the two points $(\tau_1,{f_2}(\tau_1))$ and $(\tau_2,{f_2}(\tau_2))$.
\end{lemma}
\begin{IEEEproof}
See Appendix \ref{proof:3}.
\end{IEEEproof}
Note that the two points $\tau_1$ and $\tau_2$ can be found by using the iterative computation procedure in Appendix \ref{proof:3}. Also note that $\tau_1$ should be strictly positive (though very small in general), since for Type-II symbols and at the zero spoofing power, the marginal SSER with respect to the spoofing power is negative infinity ($f_2'(0) = -\infty$).

\begin{figure}
\centering
 \epsfxsize=1\linewidth
    \includegraphics[width=8cm]{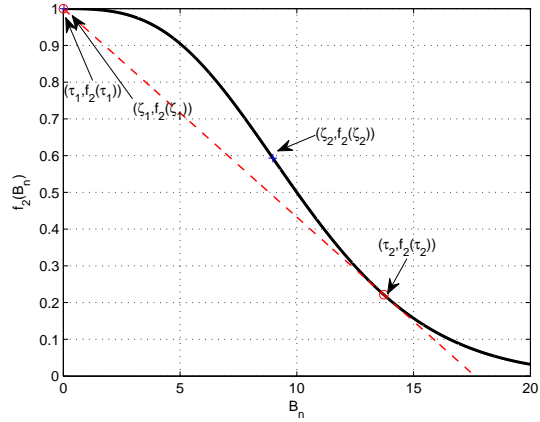}
\caption{An illustrative example of $f_2(B_n)$ with $P = 10$.} \label{fig:GBn}\vspace{-0em}
\end{figure}

For the purpose of illustration, Fig. \ref{fig:GBn} shows an example of $f_2(B_n)$ with $P = 10$, which validates the structural property of $f_2(B_n)$ in Lemmas \ref{proposition:2} and \ref{proposition:3}. It is observed that for Type-II symbols, when the spoofing power $B_n$ is between $\tau_1$ and $\tau_2$, ``time-sharing'' between the two spoofing powers $\tau_1$ and $\tau_2$ can achieve a lower SSER (or equivalently, a better spoofing performance) than using the spoofing power $B_n$ constantly.\footnote{By time-sharing, we mean that the spoofer uses the spoofing power $\tau_1$ for a $\gamma$ portion of time, and $\tau_2$ for the remaining $1 - \gamma$ portion of time, where $0\le \gamma \le1$ is uniquely chosen such that $\gamma \tau_1 + (1-\gamma)\tau_2 = B_n$ for any given $B_n>0$.} This is essential to help derive the optimal power allocation solution to problem (\ref{Problem:BPSK}), as shown next.

\begin{figure}
\centering
 \epsfxsize=1\linewidth
    \includegraphics[width=8cm]{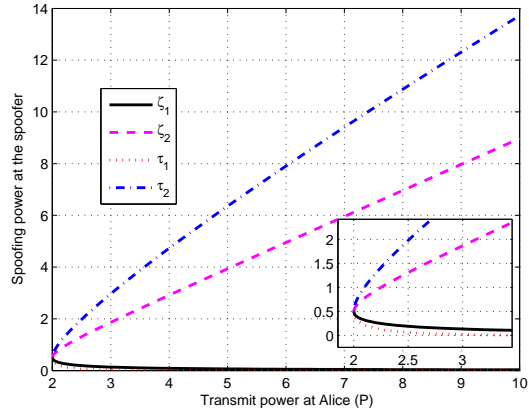}
\caption{The values of $\zeta_1$, $\zeta_2$, $\tau_1$, and $\tau_2$ under different transmit power $P$ at Alice.} \label{fig:SpoofingPower}\vspace{-2em}
\end{figure}

Furthermore, Fig. \ref{fig:SpoofingPower} shows the values of $\zeta_1$, $\zeta_2$, $\tau_1$, and $\tau_2$ versus the transmit power $P$ at Alice. It is observed that as $P$ increases, the values of $\zeta_2$ and $\tau_2$ increase while those of $\zeta_1$ and $\tau_1$ decrease. When $P > 3$, the value of $\tau_2$ is observed to be larger than $P$, while $\tau_1$ is observed to be close to zero (though strictly positive).

\subsection{Optimal Spoofing Power Allocation for Problem (\ref{Problem:BPSK})}\label{sec:spoofing:bpsk}

Now, we present the optimal solution to problem (\ref{Problem:BPSK}) by using the properties of $f_1(A_n)$ and $f_2(B_n)$ shown above. To help description, we define a new function $\bar f_2(B_n)$: when $P\le 2$, we define $\bar f_2(B_n)$ to be equivalent to $f_2(B_n)$, i.e., $\bar f_2(B_n) = f_2(B_n)$; while when $P>2$, we define
\begin{align}
\bar f_2(B_n)= \left\{ \begin{array}{ll}
f_2(B_n),&{\rm if}~B_n\in[0,\tau_1] \cup [\tau_2,+\infty) \\
cB_n + d,&{\rm if}~B_n\in(\tau_1,\tau_2),
\end{array}\right.\label{eqn:barG}
\end{align}
where $c = \frac{f_2(\tau_2)-f_2(\tau_1)}{\tau_2-\tau_1}$ and $d = f_2(\tau_2) - c\tau_2$. Here, the points $(B_n,cB_n + d)$ correspond to those on the straight line passing through the two points $(\tau_1,{f_2}(\tau_1))$ and $(\tau_2,{f_2}(\tau_2))$. Based on Lemma \ref{proposition:3}, it is evident that $\bar f_2 (B_n)$ serves as a lower bound of $f_2(B_n)$ over $B_n\in(\tau_1,\tau_2)$, and importantly, $\bar f_2(B_n)$ is convex over $B_n\in[0,+\infty)$. Accordingly, we define an auxiliary optimization problem
\begin{align}
\min_{\{A \ge 0\},\{B\ge 0\}} & \frac{1}{2}(f_1(A) + \bar f_2(B))\nonumber\\
\mathrm{s.t.}~~~~&~A+B\le 2Q, \label{Problem:BPSK:5}
\end{align}
which is convex and whose optimal solution is denoted as $A^*$ and $B^*$. Here, since the strict equality $A^* + B^* = 2Q$ should hold at the optimality of problem (\ref{Problem:BPSK:5}), $A^*$ and $B^*$ can be obtained by using a simple bisection search. Note that both $A^*$ and $B^*$ should be strictly positive, which is due to the fact that at the zero spoofing power, the marginal SSERs with respect to the spoofing power are both negative infinity ($f_1'(0) = -\infty$ and $f_2'(0) = -\infty$).

With the help of $A^*$ and $B^*$, we have the following proposition.

\begin{proposition}\label{proposition:solution}
The optimal solution of $\{A_n$\} to problem (\ref{Problem:BPSK}) is given as $A_n^* = A^*, \forall n\in\mathcal N_1,$ and that of $\{B_n\}$ is given as follows by considering two cases.
\begin{itemize}
  \item {\it When $P > 2$ and $B ^* \in (\tau_1,\tau_2)$}, the spoofer uses time-sharing between the spoofing powers $\tau_1$ and $\tau_2$, i.e., the spoofer sets $B_n^* = \tau_1$ over a $\gamma$ fraction of the symbols in $\mathcal N_2$, and $B_n^* = \tau_2$ over the remaining $1-\gamma$ fraction in $\mathcal N_2$, where $0<\gamma<1$ is uniquely chosen such that $\gamma\tau_1+(1-\gamma)\tau_2 = B^*$.
  \item {\it Otherwise}, it follows that $B_n^* = B^*, \forall n\in\mathcal N_2$.
\end{itemize}
\end{proposition}
\begin{IEEEproof}
See Appendix \ref{proof:solution}.
\end{IEEEproof}

\begin{table}[!t]\scriptsize
\caption{Algorithm for Solving Problem (\ref{Problem:BPSK})} \centering
\begin{tabular}{|p{13cm}|}
\hline\vspace{0.01cm}
1) If $P > 2$, then find the two points $\tau_1$ and $\tau_2$ by using the iterative computation procedure in Appendix \ref{proof:3}.\\
2) Construct the new function $\bar f_2(B_n)$ as in (\ref{eqn:barG}), and obtain $A^*$ and $B^*$ by solving problem (\ref{Problem:BPSK:5}).\\
3) Obtain the optimal solution $\{A_n^*\}$ and $\{B_n^*\}$ to problem (\ref{Problem:BPSK}) by Proposition \ref{proposition:solution}.\\
 \hline
\end{tabular}\label{Table:I}\vspace{0em}
\end{table}

Therefore, problem (\ref{Problem:BPSK}) is finally solved, and we summarize the algorithm to optimally solve it in Table \ref{Table:I}.

It is worth emphasizing that Proposition \ref{proposition:solution} shows the following interesting optimal spoofing power allocation strategies for the spoofer to minimize the average SSER.

\begin{itemize}
  \item When the transmit power at Alice is low (i.e., $P \le 2$) or the spoofing power at the spoofer is high (such that $B ^* > \tau_2$), the spoofer should use the optimized constant transmit power over both Type-I and Type-II symbols. This is due to the fact that both SSER functions $f_1(A_n)$ and $f_2(B_n)$ are convex over such regimes.
  \item When the transmit power $P$ at Alice is high (i.e., $P > 2$) and the spoofing power $Q$ at the spoofer is low{\footnote{Indeed, when the spoofing power is sufficiently low such that $B ^* \le \tau_1$, the spoofer should instead use constant spoofing power over Type-II symbols. Nevertheless, since $\tau_1$ is also too small, this case does not happen under practical values of $Q$.}} (such that $\tau_1 \le B ^* \le \tau_2$), the spoofer focuses its spoofing power over only a certain percentage of Type-II symbols with an ``on-off'' power control, i.e., the spoofer uses a large spoofing power (i.e., $\tau_2 > 0$) over a $1-\gamma$ portion of Type-II symbols, and uses nearly zero spoofing power over the other Type-II symbols. This is due to the fact that the SSER function $f_2(B_n)$ is non-convex over the regime of $B_n\in(\tau_1,\tau_2)$, and thus it is beneficial for the spoofer to allocate almost all the power over a limited number of Type-II symbols.
\end{itemize}


\section{Symbol-Level Spoofing Design with QPSK Signalling}\label{sec:QPSK}

In this section, we consider the case with QPSK signalling, i.e., $\mathcal{M} \triangleq \{(\pm1 \pm j)/\sqrt{2}\}$. We first design the symbol-level spoofing signals and obtain the SSER functions under any given spoofing power, and then solve the average SSER minimization problem (\ref{eqn:5}) in this case.

\subsection{Spoofing Signals Design and Problem Reformulation}

\begin{figure}
\centering
 \epsfxsize=1\linewidth
    \includegraphics[width=12cm]{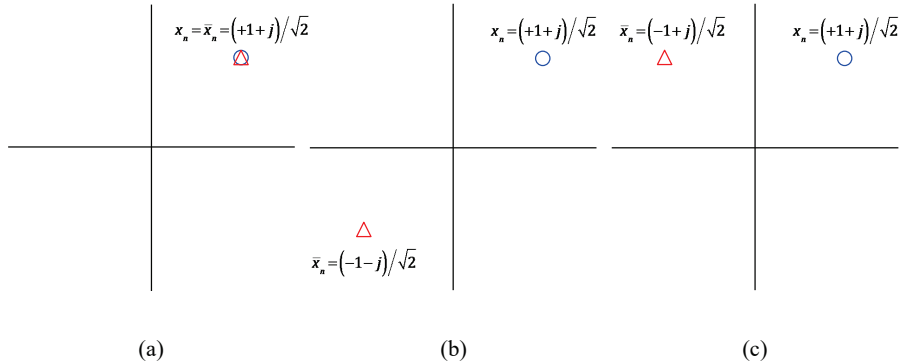}
\caption{Illustration of different types of symbols with the QPSK modulation, where the red triangle denotes the original constellation point $x_n$ of Alice, and the blue circular denotes the desirable constellation point $\bar x_n$ of the spoofer. (a) An example of Type-I symbols, where $x_n$ and $\bar x_n$ are identical with $x_n = \bar x_n = (+1+j)/\sqrt{2}$; (b) An example of Type-II symbols, where $x_n$ and $\bar x_n$ are opposite with $x_n = (+1+j)/\sqrt{2}$ and $\bar x_n = (-1-j)/\sqrt{2}$; (c) An example of Type-III symbols, where $x_n$ and $\bar x_n$ are neighboring with $x_n = (+1+j)/\sqrt{2}$ and $\bar x_n = (-1+j)/\sqrt{2}$.} \label{fig:QPSK}\vspace{-0em}
\end{figure}

Similar to the BPSK case and as illustrated in the example in Fig. \ref{fig:QPSK}, we classify the QPSK symbols into three types based on the relationship between the original constellation point $x_n$ of Alice and the desirable one $\bar x_n$ of the spoofer.
\begin{itemize}
  \item {\it Type-I symbol:} The symbol $n$ is called a Type-I symbol if $x_n $ and $\bar x_n$ are identical (i.e., $x_n = \bar x_n$). The set of all Type-I symbols is denoted as $\mathcal N_1$.
  \item {\it Type-II symbol:} The symbol $n$ is called a Type-II symbol if $x_n $ and $\bar x_n$ are opposite (i.e., $x_n = - \bar x_n$). The set of all Type-II symbols is denoted as $\mathcal N_2$.
  \item {\it Type-III symbol:} The symbol $n$ is called a Type-III symbol if $x_n $ and $\bar x_n$ are neighboring. The set of all Type-III symbols is denoted as $\mathcal N_3$.
\end{itemize}
Here, Type-I, Type-II, and Type-III symbols on average occupy $1/4$, $1/4$, and $1/2$ portions of all symbols, respectively. To facilitate the description, we focus on one particular original constellation point $x_n = (1+j)/\sqrt{2}$, and consider the desirable constellation point to be $\bar x_n = (1+j)/\sqrt{2}$, $\bar x_n = (-1-j)/\sqrt{2}$, and $\bar x_n = (-1+j)/\sqrt{2}$, for Type-I, Type-II, and Type-III symbols, respectively. Under each of the three desirable constellation points, we will design the corresponding symbol-level spoofing signal and derive the SSER function under any given spoofing power. Note that the spoofing signals design for other symbols (i.e., Type-I, Type-II, and Type-III symbols other than those in Fig. \ref{fig:QPSK}) can be similarly devised to achieve the same SSER functions, and thus is omitted for brevity.

First, consider a particular Type-I symbol $n\in\mathcal N_1$ with $x_n = \bar x_n = (1+j)/\sqrt{2}$. In this case, the optimal spoofing signal is given in the following proposition.
\begin{proposition}\label{proposition:QPSK:1}
It is optimal to minimize the conditional SSER ${\rm Pr}(\hat x_n\neq \bar x_n|x_n = \bar x_n =  (1+j)/\sqrt{2})$ by designing $z_n = \sqrt{A_n} (1+j)/\sqrt{2})$, where $A_n$ denotes the given spoofing power for this Type-I symbol. Accordingly, ${\rm Pr}(\hat x_n\neq \bar x_n|x_n = \bar x_n =  (1+j)/\sqrt{2})$ is given as
\begin{align}\label{eqn:tilde:f1}
g_1(A_n) = & 1- \left(\frac{1}{2}+\frac{1}{2}{\rm{erf}}\big(\sqrt{A_n/2}+\sqrt{P/2}\big)\right)^2.
\end{align}
\end{proposition}
\begin{IEEEproof}
See Appendix \ref{Appendix:E}.
\end{IEEEproof}

Next, consider a particular Type-II symbol $n\in\mathcal N_2$ with $x_n = (1+j)/\sqrt{2}$ and $\bar x_n = (-1-j)/\sqrt{2}$. In this case, it is difficult to rigorously derive the optimal spoofing signal design under any values of $P$. Nevertheless, we can provide the optimal spoofing signal in the special case of $P \le 4$ in the following proposition.
\begin{proposition}\label{proposition:QPSK:2}
In the case of $P \le 4$, it is optimal to minimize the conditional SSER ${\rm Pr}(\hat x_n\neq \bar x_n|x_n = (1+j)/\sqrt{2},\bar x_n =  (-1-j)/\sqrt{2})$ by designing $z_n = \sqrt{B_n} (-1-j)/\sqrt{2}$, where $B_n$ denotes the given spoofing power for this Type-II symbol. Accordingly, ${\rm Pr}(\hat x_n\neq \bar x_n|x_n = (1+j)/\sqrt{2},\bar x_n =  (-1-j)/\sqrt{2})$ is given as
\begin{align}\label{eqn:tilde:f2}
g_2(B_n) = 1- \left(\frac{1}{2}+\frac{1}{2}{\rm{erf}}\big(\sqrt{B_n/2}-\sqrt{P/2}\big)\right)^2.
\end{align}
\end{proposition}
\begin{IEEEproof}
See Appendix \ref{Appendix:F}.
\end{IEEEproof}
For the remaining case of $P > 4$, it is difficult to prove the optimality of the spoofing signal design of $z_n = \sqrt{B_n} (-1-j)/\sqrt{2}$. Nevertheless, such optimality is observed via extensive simulations. Therefore, we choose $z_n = \sqrt{B_n} (-1-j)/\sqrt{2}$ for this particular Type-II symbol under any value of $P$, and accordingly, we have the conditional SSER as $g_2(B_n)$ in (\ref{eqn:tilde:f2}).

\begin{remark}
From Propositions \ref{proposition:QPSK:1} and \ref{proposition:QPSK:2}, it is observed that the optimally designed spoofing signals for Type-I and Type-II symbols have an equal strength in their respective real and imaginary components, such that at the receiver of Bob they are constructively and destructively combined with the original signals of Alice, respectively. The design of spoofing signals in Type-I and Type-II symbols in the QPSK case is similar to that in the BPSK case (see Propositions \ref{proposition:TypeI} and \ref{proposition:TypeII}), but leading to different SSER functions due to their difference in the modulation order.
\end{remark}

In addition, consider a particular Type-III symbol $n\in\mathcal N_3$ with $x_n = (1+j)/\sqrt{2}$ and $\bar x_n = (-1+j)/\sqrt{2}$. In this case, we independently design the real and imaginary components of the spoofing signal, and generally set it to be $z_n = -\sqrt{C^{\rm R}_n} + j\sqrt{C^{\rm I}_n}$, where the spoofing power is denoted as $C^{\rm R}_n + C^{\rm I}_n$. Under such a design, the conditional SSER is expressed as
\begin{align}\label{eqn:g3:000}
g_3(C^{\rm R}_n,C^{\rm I}_n) = & 1- \left(\frac{1}{2}+\frac{1}{2}{\rm{erf}}\big(\sqrt{C^{\rm R}_n}-\sqrt{P/2}\big)\right)\left(\frac{1}{2}+\frac{1}{2}{\rm{erf}}\big(\sqrt{C^{\rm I}_n}+\sqrt{P/2}\big)\right).
\end{align}
Here, the derivation of (\ref{eqn:g3:000}) is based on a similar procedure as in the proof of Propositions \ref{proposition:QPSK:1} (see (\ref{eqn:g1})), and thus is omitted for brevity.

%

By combining the above three types of symbols,
the average SSER minimization problem is reformulated as a spoofing power allocation problem among the three types of symbols, given as
\begin{align}
\min_{\{A_n\ge 0\},\{B_n\ge 0\},\{C_n^{\rm R}\ge 0,C_n^{\rm I}\ge 0\}}~&\frac{1}{4}\mathbb{E}_{n\in\mathcal N_1}(g_1(A_n))
+\frac{1}{4}\mathbb{E}_{n\in\mathcal N_2}(g_2(B_n))
+\frac{1}{2}\mathbb{E}_{n\in\mathcal N_3}(g_3(C^{\rm R}_n,C^{\rm I}_n)) \nonumber\\
\mathrm{s.t.}~~~~~&\frac{1}{4}\mathbb{E}_{n\in\mathcal N_1}(A_n)+\frac{1}{4}\mathbb{E}_{n\in\mathcal N_2}(B_n) +\frac{1}{2}\mathbb{E}_{n\in\mathcal N_3}(C^{\rm R}_n+C^{\rm I}_n) \le Q.\label{eqn:problem:QPSK}
\end{align}
Problem (\ref{eqn:problem:QPSK}) is nonconvex in general and thus difficult to solve. In the following, we show some useful properties of the three SSER functions, to help solve problem (\ref{eqn:problem:QPSK}).

\subsection{Properties of the SSER Functions $g_1(A_n)$, $g_2(B_n)$, and $g_3(C^{\rm R}_n,C^{\rm I}_n)$}

In this subsection, we show the monotonic properties and convexities of the three SSER functions.

\begin{lemma}\label{lemma:QPSK:1}
$g_1(A_n)$ is monotonically decreasing and convex over $A_n \in [0,+\infty)$.
\end{lemma}
\begin{IEEEproof}
See Appendix \ref{Appendix:G}.
\end{IEEEproof}

For $g_2(B_n)$, it is very difficult for us to rigorously prove its convexity over the whole regime of $B_n \in [0,+\infty]$. We first provide the following lemma to analytically show its convexity under certain regimes, and then remark on its convexity in the general case.

\begin{lemma}\label{lemma:QPSK:2}
$g_2(B_n)$ is monotonically decreasing over $B_n \in [0,+\infty)$. The convexity of $g_2(B_n)$ is given as follows.
\begin{itemize}
  \item Under any value of $P$, there exists a small but positive $\chi_1$, such that $g_2(B_n)$ is convex over $B_n \in [0,\chi_1]$, where $\chi_1 < \zeta_1$ with $\zeta_1$ given in (\ref{eqn:zeta1});
  \item Under any value of $P$, $g_2(B_n)$ is convex over $B_n \in [\chi_2,+\infty)$, where $\chi_2$ is given as follows and $\chi_2 > \zeta_2$ with $\zeta_2$ given in (\ref{eqn:zeta2});
  \begin{align}
  \chi_2 = \max\left(P,\left(\frac{\sqrt{P} + \sqrt{\frac{\pi}{2}} + \sqrt{\left(\sqrt{P} + \sqrt{\frac{\pi}{2}}\right)^2-2}}{2}\right)^2\right);
  \end{align}
  \item When $P > 2$, $g_2(B_n)$ is concave over $B_n \in[\zeta_1,\zeta_2]$.
\end{itemize}
\end{lemma}
\begin{IEEEproof}
See Appendix \ref{Appendix:H}.
\end{IEEEproof}

\begin{remark}\label{remark:g2B}
Note that in Lemma \ref{lemma:QPSK:2}, we cannot analytically show the convexity of $g_2(B_n)$ in the regime of $B_n \in (\zeta_1,\chi_1) \cup (\zeta_2,\chi_2)$, and thus in the whole regime of $B_n \in [0,+\infty)$. Despite this fact, via extensive simulations, we numerically find that $g_2(B_n)$ has a similar convexity property as $f_2(B_n)$ in Proposition \ref{proposition:2}. That is, under Alice's low transmit power regime (particularly, when $P$ is no larger than a boundary point  $\xi \approx1.146$), $g_2(B_n)$ is convex over $B_n \in [0,+\infty)$; whereas under Alice's high transmit power regime (when $P > \xi \approx1.146$), there exist two points $0\le \bar \zeta_1 \le \bar \zeta_2$ such that $g_2(B_n)$ is first convex over $[0,\bar\zeta_1]$, then concave over $(\bar\zeta_1,\bar\zeta_2)$, and finally convex over $[\bar\zeta_2,+\infty)$. In the latter case, it follows similar to Lemma \ref{proposition:3} that there exist two points $\bar \tau_1$ and $\bar\tau_2$ with $0 < \bar\tau_1 \le \bar\zeta_1$ and $\bar\tau_2 \ge \bar\zeta_2$, such that all the points $(B_n,g_2(B_n))$ are above the straight line passing through the two points $(\bar\tau_1,{g_2}(\bar\tau_1))$ and $(\bar\tau_2,{g_2}(\bar\tau_2))$. Note that under any given value of $P$, the values of $\bar\zeta_1$ and $\bar\zeta_2$ can be numerically found by checking the second-order derivatives of $g_2(B_n)$; and baed on them we can obtain $\bar\tau_1$ and $\bar\tau_2$ by using a similar procedure as that in Appendix \ref{proof:3}.
\end{remark}


Next, we consider the SSER function $g_3(C_n^{\rm R},C_n^{\rm I})$ for the Type-III symbols. We rewrite $g_3(C_n^{\rm R},C_n^{\rm I}) = 1-g_3^{\rm R}(C_n^{\rm R})g_3^{\rm I}(C_n^{\rm I})$ with $g_3^{\rm R}(C_n^{\rm R})=  \frac{1}{2}+\frac{1}{2}{\rm{erf}}\big(\sqrt{C^{\rm R}_n}-\sqrt{P/2}\big)$ and $g_3^{\rm I}(C_n^{\rm I}) =\frac{1}{2}+\frac{1}{2}{\rm{erf}}\big(\sqrt{C^{\rm I}_n}+\sqrt{P/2}\big)$. Then we have the following lemma.
\begin{lemma}\label{lemma:QPSK:3}
$g_3^{\rm I}(C_n^{\rm I})$ is monotonically increasing and concave over $C_n^{\rm I} \in [0,+\infty)$. $g_3^{\rm R}(C_n^{\rm R})$ is monotonically increasing over $C_n^{\rm R} \in [0,+\infty)$. The convexity of $g_3^{\rm R}(C_n^{\rm R})$ is given as follows depending on Alice's transmit power $P$.
\begin{itemize}
  \item {\it Alice's low transmit power regime (i.e., $P \le 4$)}: $g_3^{\rm R}(C_n^{\rm R})$ is concave over $B_n \in [0,+\infty)$.
  \item {\it Alice's high transmit power regime (i.e., $P > 4$)}: $g_3^{\rm R}(C_n^{\rm R})$ is first concave over $B_n \in [0,\hat\zeta_1]$, then convex over $B_n \in(\hat\zeta_1,\hat\zeta_2)$, and finally concave over $B_n\in[\hat\zeta_2,+\infty)$, where the two boundary points $\hat\zeta_1 < \hat\zeta_2$ are given as $\hat\zeta_1 = \left(\frac{\sqrt{P/2}-\sqrt{P/2-2}}{2}\right)^2$ and $\hat\zeta_2 = \left(\frac{\sqrt{P/2}+\sqrt{P/2-2}}{2}\right)^2$. Furthermore, there exist two points $\hat\tau_1$ and $\hat\tau_2$ with $0 < \hat\tau_1 \le \hat\zeta_1$ and $\hat\tau_2 \ge \hat\zeta_2$, such that all the points $g_3^{\rm R}(C_n^{\rm R})$'s are below the straight line passing through the two points $(\hat\tau_1,g_3^{\rm R}(\hat\tau_1))$ and $(\hat\tau_2,g_3^{\rm R}(\hat\tau_2))$.
\end{itemize}
\end{lemma}
\begin{IEEEproof}
This lemma can be proved following similar procedures as those for Lemmas \ref{proposition:1} and \ref{proposition:2}. Therefore, the details are omitted for brevity.
\end{IEEEproof}

The results in Lemmas \ref{lemma:QPSK:2} and \ref{lemma:QPSK:3} will play important roles in the design of the spoofing power allocation to solve problem (\ref{eqn:problem:QPSK}), as will be shown next.




\subsection{Spoofing Power Allocation for Problem (\ref{eqn:problem:QPSK})}

In this subsection, we propose the optimal solution to problem (\ref{eqn:problem:QPSK}) by using the properties of the SSER functions shown in the proceeding subsection. First, we define two auxiliary SSER functions for Type-II and Type-III symbols to facilitate the derivation. For Type-II symbols, we define an auxiliary SSER function $\bar g_2(B_n)$, where if $P \le \xi \approx 1.146$, we have $\bar g_2(B_n) = g_2(B_n), \forall B_n \in[0,+\infty)$; whereas if $P > \xi \approx1.146$, it follows that
\begin{align}\label{eqn:bar:g2}
\bar g_2(B_n) =\left\{
\begin{array}{ll}
g_2(B_n), &{\rm if}~ B_n \in [0,\bar\tau_1]\cup[\bar\tau_2, +\infty) \\
\bar c B_n + \bar d, & {\rm if}~ B_n \in (\bar\tau_1,\bar\tau_2),
\end{array}
\right.
\end{align}
where $\bar c = (g_2(\bar\tau_2) - g_2(\bar\tau_1))/(\bar\tau_2  -\bar\tau_1)$ and $\bar d = g_2(\bar\tau_2) - \bar c \bar\tau_2$. Here, the points $(B_n,\bar cB_n +\bar d)$ correspond to those on the straight line passing through the two points $(\bar\tau_1,{g_2}(\bar\tau_1))$ and $(\bar\tau_2,{g_2}(\bar\tau_2))$. Based on Lemma \ref{proposition:3}, it is evident that $\bar g_2 (B_n)$ serves as a lower bound of $ g_2(B_n)$ over $B_n\in(\bar\tau_1,\bar\tau_2)$, and importantly, $\bar g_2(B_n)$ is convex over $B_n\in[0,+\infty)$.


In addition, we consider Type-III symbols, and define another auxiliary function
\begin{align}
\bar g_3(C_n^{\rm R},C_n^{\rm I}) =1-\bar g_3^{\rm R}(C_n^{\rm R})g_3^{\rm I}(C_n^{\rm I}),
\end{align}
where if $P\le4$, we have $\bar g^{\rm R}_3(C_n^{\rm R}) = g^{\rm R}_3(C_n^{\rm R})$; whereas if $P > 4$, it follows that
\begin{align}\label{eqn:bar:g3}
\bar g^{\rm R}_3(C_n^{\rm R}) =\left\{
\begin{array}{ll}
g_3^{\rm R}(C_n^{\rm R}), &{\rm if}~ C_n^{\rm R} \in [0,\hat\tau_1] \cup [\hat\tau_2,+\infty) \\
\hat c C_n^{\rm R}+ \hat d, & {\rm if}~C_n^{\rm R} \in (\hat\tau_1,\hat\tau_2),
\end{array}
\right.
\end{align}
with $\hat c = \frac{g_3^{\rm R}(\hat\tau_2)-g_3^{\rm R}(\hat\tau_1)}{\hat\tau_2-\hat\tau_1}$ and $\hat d = g_2(\hat\tau_2) - \hat c\hat\tau_2$. Here, the points $(C_n^{\rm R},\hat c C_n^{\rm R} +\hat d)$ correspond to those on the straight line passing through the two points $(\hat\tau_1,{g_3}^{\rm R}(\hat\tau_1))$ and $(\hat\tau_2,{g_3}^{\rm R}(\hat\tau_2))$. Based on Lemma \ref{lemma:QPSK:3}, it is evident that $\bar g_3^{\rm R} (C_n^{\rm R})$ serves as an upper bound of $ g_3^{\rm R}(C_n^{\rm R})$ over $C_n^{\rm R}\in(\hat\tau_1,\hat\tau_2)$, and accordingly, $\bar g_3(C_n^{\rm R},C_n^{\rm I})$ serves as a lower bound of $g_3(C_n^{\rm R},C_n^{\rm I})$ over $C_n^{\rm R}\in(\hat\tau_1,\hat\tau_2)$. Furthermore, $\bar g_3(C_n^{\rm R})$ is concave over $C_n^{\rm R}\in[0,+\infty)$.

By combining the above discussions for the three types of symbols, we solve problem (\ref{eqn:problem:QPSK}) by solving the following auxiliary problem:
\begin{align}
\min_{A\ge 0,B\ge 0,C^{\rm R}\ge0,C^{\rm I}\ge0}~&\frac{1}{4}\bar g_1(A) +\frac{1}{4}\bar g_2(B) +\frac{1}{2}\left(1-\bar g_3^{\rm R}(C^{\rm R})g_3^{\rm I}(C^{\rm I})\right) \nonumber\\
\mathrm{s.t.}~~~~~&\frac{1}{4}A+\frac{1}{4} B +\frac{1}{2}(C^{\rm R}+C^{\rm I}) \le Q.\label{eqn:problem:QPSK:3}
\end{align}

Note that problem (\ref{eqn:problem:QPSK:3}) itself is non-convex due to the coupling of $\bar g_3^{\rm R}(C^{\rm R})$ and $g_3^{\rm I}(C^{\rm I})$. Nevertheless, under any given $C^{\rm R} \ge 0$, the optimization over $A$, $B$, and $C^{\rm I}$ becomes a convex optimization problem. As a result, we use a one-dimensional search over $C^{\rm R} \in [0,2Q]$, and solve the convex optimization problem in (\ref{eqn:problem:QPSK:3}) under any given $C^{\rm R} $ to obtain the optimal $A$, $B$, and $C^{\rm I}$. Therefore, problem (\ref{eqn:problem:QPSK:3}) is optimally solved, for which the corresponding spoofing power allocation solution is denoted as $A^{**}$, $B^{**}$, $C^{{\rm R}**}$, and $C^{{\rm I}**}$, respectively. Then we obtain the optimal spoofing signals design for problem (\ref{eqn:problem:QPSK}) as given in the following proposition.


\begin{proposition}\label{proposition:4.3}
The spoofing power allocation solution of $\{A_n\}$ and $\{C_n^{\rm I}\}$ to problem (\ref{eqn:problem:QPSK}) is given as $A_n^{**} = A^{**}, \forall n\in\mathcal N_1,$ and $C_n^{{\rm I}**} = C^{{\rm I}**}, \forall n\in\mathcal N_3$, and that of $\{B_n\}$ and $\{C_n^{\rm R}\}$ is given as follows.
\begin{itemize}
  \item {\it When $P > \xi \approx 1.146$ and $B ^{**} \in (\bar \tau_1,\bar\tau_2)$}, the spoofer uses time-sharing between the spoofing power $\bar \tau_1$ and $\bar \tau_2$, i.e., the spoofer sets $B_n^* = \bar \tau_1$ over a $\bar\gamma$ fraction of the symbols in $\mathcal N_2$, and $B_n^* = \bar \tau_2$ over the remaining $1-\bar\gamma$ fraction in $\mathcal N_2$, where $0<\bar\gamma<1$ is uniquely chosen such that $\bar\gamma\bar \tau_1 + (1-\bar\gamma)\bar \tau_2 = B^{**}$; {\it otherwise}, it follows that $B_n^{**} = B^{**}, \forall n\in\mathcal N_2$.
  \item {\it When $P > 4$ and $C ^{\rm R**} \in (\hat \tau_1,\hat\tau_2)$}, the spoofer uses time-sharing between the spoofing power $\hat\tau_1$ and $\hat\tau_2$, i.e., the spoofer sets $C_n^{\rm R**} = \hat\tau_1$ over a $\hat\gamma$ fraction of the symbols in $\mathcal N_2$, and $C_n^{\rm R**} = \hat\tau_2$ over the remaining $1-\hat\gamma$ fraction in $\mathcal N_2$, where $0<\hat\gamma<1$ is uniquely chosen such that $\hat\gamma\hat\tau_1+(1-\hat\gamma)\hat\tau_2 = C ^{\rm R**}$; {\it otherwise}, it follows that $C_n^{\rm R**} = C ^{\rm R**}, \forall n\in\mathcal N_3$.
\end{itemize}
\end{proposition}

\begin{IEEEproof}
This proposition can be proved following similar procedures as that for Proposition \ref{proposition:solution}. Therefore, the details are omitted for brevity.
\end{IEEEproof}

\begin{table}[!t]\scriptsize
\caption{Algorithm for Solving Problem (\ref{eqn:problem:QPSK})} \centering
\begin{tabular}{|p{13cm}|}
\hline\vspace{0.01cm}
1) If $P > \xi \approx 1.146$, then find the two points $\bar\tau_1$ and $\bar\tau_2$, and construct the new function $\bar g_2(B_n)$ as in (\ref{eqn:bar:g2}).\\
2) If $P > 4$, then find the two points $\hat\tau_1$ and $\hat\tau_2$, and construct the new function $\bar g^{\rm R}_3(C_n^{\rm R})$ as in (\ref{eqn:bar:g3}).\\
3) Obtain the optimal solution to problem (\ref{eqn:problem:QPSK:3}) to be $A^{**}$, $B^{**}$, $C^{{\rm R}**}$, and $C^{{\rm I}**}$.\\
4) Obtain the optimal solution $\{A_n^{**}\}$, $\{B_n^{**}\}$, $\{C_n^{{\rm R}**}\}$, and $\{C_n^{{\rm I}**}\}$ to problem (\ref{eqn:problem:QPSK}) by Proposition \ref{proposition:4.3}.\\
 \hline
\end{tabular}\label{Table:II}\vspace{0em}
\end{table}

Therefore, problem (\ref{eqn:problem:QPSK}) is finally solved, and we summarize the algorithm to solve it in Table \ref{Table:II}.

\section{Numerical Results}\label{sec:4}

In this section, we present numerical results to show the performance of our proposed symbol-level spoofing with optimized power allocation, as compared with two benchmark schemes in the following.
\begin{itemize}


\item {\it Block-level spoofing:} In this scheme, the spoofer is assumed to be not aware of the original symbol information $x_n$ from Alice. In this scheme, the spoofer uses the constant transmit power $Q$ over all symbols{\footnote{Due to the non-convexity of the SSER function, it is possible to further improve the average SSER performance of the block-level spoofing by allowing adaptive power allocation over symbols (e.g., time-sharing of various spoofing powers). Nevertheless, how to optimize the adaptive power allocation is a non-trivial problem, which is left for our future work.}} and sets the spoofing signal to be the exact desirable constellation point $\bar x_n$, i.e., $z_n = \sqrt{Q}\bar x_n $.

\item {\it Heuristic symbol-level spoofing:} In this scheme, the spoofer designs its spoofing signals by only heuristically exploiting the symbol-level relationship between the original constellation points of Alice and the desirable one of the spoofer, but without the sophisticated transmit optimization as in our proposed optimal symbol-level spoofing. In particular, for Type-I symbols, the spoofer does not allocate any spoofing power to them, since the original and desirable constellation points are already identical; for other symbols (i.e., Type-II symbols for the BPSK case, as well as Type-II and Type-III symbols for the QPSK case), the spoofer equally allocates its spoofing power to each of them. As a result, in the BPSK case, the spoofing powers allocated for each Type-I and Type-II symbols are $0$ and $2Q$, respectively, and the resultant average SSER is given as $\frac{1}{2}f_1(0) +  \frac{1}{2} f_2(2Q)$. In the QPSK case, the spoofing powers allocated for each Type-I, Type-II, and Type-III symbols are $0$, $4Q/3$, and $4Q/3$, respectively, and the resultant average SSER is given to be $\frac{1}{4}g_1(0)  + \frac{1}{4}g_2(4Q/3) +   \frac{1}{2}g_3(4Q/3,0)$. Here, the average SSER $\frac{1}{2}g_3(4Q/3,0)$ for Type-III symbols is obtained by considering the symbols with $x_n = (1+j)/\sqrt{2}$ and $\bar x_n = (-1+j)/\sqrt{2}$, and allocating the spoofing power to the real components only (as the imaginary components of the original and desirable constellation points are already identical).

\end{itemize}

\begin{figure}
\centering
 \epsfxsize=1\linewidth
    \includegraphics[width=8cm]{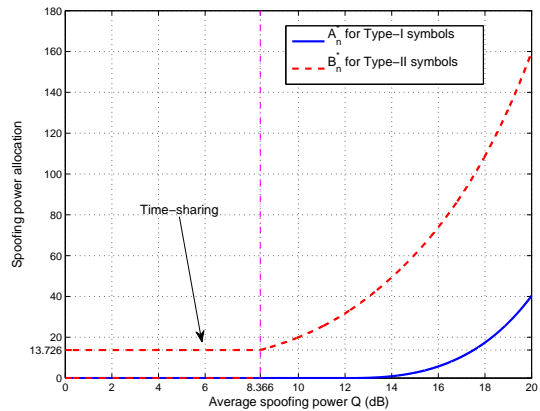}
\caption{The optimal spoofing power allocation versus the average spoofing power $Q$, where the transmit power at Alice is set as $P = 10$.} \label{fig:4}\vspace{-0em}
\end{figure}

First, consider the BPSK case, and Fig. \ref{fig:4} shows the optimal spoofing power allocation versus the average spoofing power $Q$, where the transmit power at Alice is set as $P = 10$. It is observed that when $Q$ is small (i.e., $Q \le 13$ dB), almost all the spoofing power is reserved for Type-II symbols to move the constellation points efficiently towards the desirable opposite directions; whereas when $Q$ becomes large (i.e., $Q > 13$ dB), the spoofing power is allocated more fairly between Type-I and Type-II symbols. Particularly, in the small $Q$ regime when $Q < 8.366$ dB, it is observed that an ``on-off'' time-sharing strategy between the spoofing power $\tau_1 \approx 0$ and $\tau_2 = 13.726$ should be employed over Type-II symbols. In other words, in this regime, the spoofer should focus its spoofing power on a certain portion of Type-II symbols.

\begin{figure}
\centering
 \epsfxsize=1\linewidth
    \includegraphics[width=8cm]{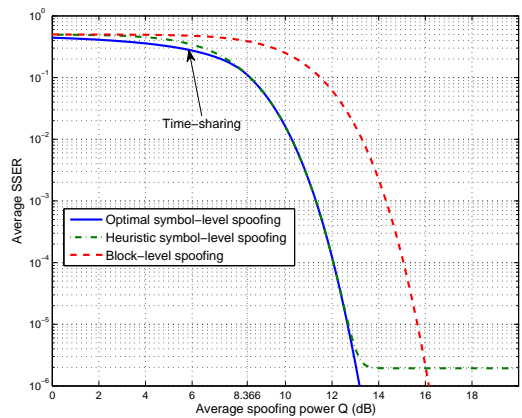}
\caption{The average SSER performance in the BPSK case, where the transmit power at Alice is set as $P = 10$.} \label{fig:2}\vspace{-2em}
\end{figure}

Fig. \ref{fig:2} shows the average SSER performance of the three schemes versus the spoofing power $Q$ in the BPSK case, where the transmit power at Alice is set as $P = 10$. It is observed that the optimal symbol-level spoofing achieves a better performance (or equivalently, a lower average SSER) than the block-level spoofing benchmark. In particular, over 3 dB performance gain is obtained by the symbol-level spoofing when the average spoofing power $Q$ becomes large. It is also observed that the optimal symbol-level spoofing leads to a lower average SSER than the heuristic symbol-level spoofing when $Q < 8.366$ dB and $Q > 13$ dB, and the two schemes have a similar average SSER performance when the value of $Q$ is between 8.366 dB and 13 dB. The results can be explained based on the optimal spoofing power allocation shown in Fig. \ref{fig:4}. When $Q < 8.366$ dB, the optimal symbol-level spoofing employs an ``on-off'' transmission strategy with time-sharing between the spoofing power $\tau_1 \approx 0$ and $\tau_2 = 13.726$, thus outperforming the heuristic one that uses fixed spoofing power over all Type-II symbols. When $Q > 13$ dB, the optimal symbol-level spoofing allocates the spoofing power more fairly between Type-I and Type-II symbols, thus reducing the average SSER as compared to the heuristic one that only allocates the spoofing power to Type-II symbols. These results show the significance of our proposed optimal spoofing power allocation. In addition, it is observed that the heuristic symbol-level spoofing performs worse than the block-level spoofing when $Q$ becomes large, which is due to the fact that the heuristic symbol-level spoofing does not allocate any spoofing power to the Type-I symbols, which leads to the average SSER floor.

\begin{figure}
\centering
 \epsfxsize=1\linewidth
    \includegraphics[width=8cm]{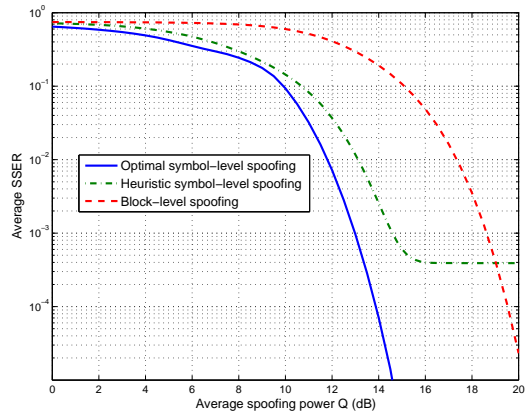}
\caption{The average SSER performance in the QPSK case, where the transmit power at Alice is set as $P = 10$.} \label{fig:222}\vspace{-2em}
\end{figure}

Next, consider the QPSK case. Fig. \ref{fig:222} shows the average SSER achieved by the three schemes versus the average spoofing power $Q$, where the transmit power at Alice is set as $P = 10$. Similar to the BPSK case as in Fig. \ref{fig:2}, the optimal symbol-level spoofing is observed to achieve significantly lower SSER than the block-level spoofing, and over 5 dB average SSER reduction is obtained when $Q$ becomes large. Furthermore, the optimal symbol-level spoofing achieves lower average SSER than the heuristic symbol-level spoofing under any value of $Q$. Based on these observations, it follows that the optimal spoofing power allocation is more important with higher-order modulations.

\section{Conclusion}\label{sec:5}

This paper proposed spoofing attacks in the physical layer for the legitimate intervention of malicious wireless communications. We proposed a new symbol-level spoofing approach for a legitimate spoofer to change the messages transmitted in a malicious link. With knowledge of the original constellation points by Alice, the spoofer exploits the correlations between its desirable constellation points and the original ones by Alice to improve the spoofing performance. In particular, we developed optimal spoofing signals design and power allocation in the cases of BPSK and QPSK modulations. How to extend the symbol-level spoofing into general modulation techniques (such as $M$-PSK and $M$-QAM modulations with $M>4$) and practical cases with fading channels and imperfect/partial transmitted message knowledge are interesting problems worth pursuing in the future.

\appendix

\subsection{Proof of Proposition \ref{proposition:TypeI}}\label{proof:TypeI}

Consider a typical Type-I symbol $n\in\mathcal N_1$ with $x_n = \bar x_n$. In this case, the SSER $\mathrm{Pr}(\hat x_n \neq \bar x_n|x_n = \bar x_n)$ is expressed as
\begin{align}
&\mathrm{Pr}(\hat x_n \neq \bar x_n|x_n = \bar x_n) = \mathrm{Pr}(\hat x_n \neq x_n) \nonumber\\
=&\mathrm{Pr}(|y_n - \sqrt{P} x_n|^2 > |y_n + \sqrt{P} x_n|^2)\label{eqn:6}\\
=&\mathrm{Pr}(|z_n+v_n|^2 > |2 \sqrt{P}x_n + z_n+v_n|^2)\label{eqn:6:1},
\end{align}
where (\ref{eqn:6}) follows from \cite[Chapter 5]{GlodsmithBook} based on the ML detection in (\ref{eqn:hatx_n}), and (\ref{eqn:6:1}) holds from (\ref{eqn:y_n}). In this case, to minimize the SSER in (\ref{eqn:6:1}), we should set the phase of $z_n$ to be same as that of $x_n$, and accordingly we have $z_n =  \sqrt{A_n}x_n$. In this case, note that the term in (\ref{eqn:6:1}) is only dependent on the real part of the CSCG random variable $v_n$, which is a real Gaussian random variable denoted by $\bar v_n$ with zero mean and variance $1/2$. Therefore, we further express the function $\mathrm{Pr}(\hat x_n \neq \bar x_n|x_n = \bar x_n)$ as
\begin{align}
f_1(A_n)=&\mathrm{Pr}(|\sqrt{A_n}x_n +\bar v_n|^2 > |2 \sqrt{P}x_n + \sqrt{A_n}x_n+\bar v_n|^2)\nonumber \\
=&\mathrm{Pr}\left(\bar v_n/x_n > \sqrt{A_n} + \sqrt{P} \right)\nonumber\\
=&\frac{1}{2} - \frac{1}{2}\mathrm{erf}\left(\sqrt{A_n}+ \sqrt{P}\right),\label{eqn:8}
\end{align}
where (\ref{eqn:8}) holds due to the fact that $\bar v_n/x_n$ is also a real Gaussian random variable with zero mean and variance $1/2$. Therefore, this proposition is proved.

\subsection{Proof of Lemma \ref{proposition:2}}\label{proof:2}
First, the first-order derivative of $f_2(B_n)$ is given as
\begin{align*}
f_2'(B_n)& = -\frac{1}{2\sqrt{\pi}} e^{-\left( \sqrt{B_n}-\sqrt{P}\right)^2}B_n^{-\frac{1}{2}} \le 0.
\end{align*}
Therefore, $f_2(B_n)$ is monotonically decreasing over $B_n \in [0,+\infty)$.

Next, we have the second-order derivative of $f_2(B_n)$ as
\begin{align*}
f_2''(B_n)& = \frac{B_n^{-\frac32}}{{4\sqrt{\pi}}} e^{-\left(\sqrt{B_n}-\sqrt{P}\right)^2} \left( 2B_n - 2 \sqrt{PB_n}+ 1
\right),
\end{align*}
based on which we consider the following two cases when $P \le 2$ and $P > 2$, respectively.
\begin{itemize}
  \item In the Alice's low transmit power regime (i.e., $P \le 2$), it can be shown that $f_2''(B_n) \ge 0$ always holds, and therefore, $f_2(B_n)$ is convex over $B_n \in [0,+\infty)$.
  \item In the Alice's high transmit power regime (i.e., $P > 2$), the equation $f_2''(B_n) = 0$ (equivalently, $2B_n - 2 \sqrt{PB_n}+ 1=0$) has two solutions given as $\zeta_1$ and $\zeta_2$ in (\ref{eqn:zeta1}) and (\ref{eqn:zeta2}), respectively. It is evident that $f_2''(B_n) \ge 0$ when $B_n \in [0,\zeta_1]$ and $B_n\in[\zeta_2,+\infty)$, and $f_2''(B_n) < 0$  when $B_n \in(\zeta_1,\zeta_2)$. Therefore, it follows that $f_2(B_n)$ is convex over $B_n \in [0,\zeta_1]$, concave over $B_n \in(\zeta_1,\zeta_2)$, and convex over $B_n\in[\zeta_2,+\infty)$.
\end{itemize}

As a result, this lemma is proved.

\subsection{Proof of Lemma \ref{proposition:3}}\label{proof:3}

We prove this lemma via two steps. First, we find two points $(\tau_1,f_2(\tau_1))$ and $(\tau_2,f_2(\tau_2))$ with $0 < \tau_1 \le \zeta_1$ and $\tau_2 \ge \zeta_2$, which satisfy $f_2'(\tau_1) = f_2'(\tau_2) = \theta(\tau_1,\tau_2)$. Here, $\theta(\tau_1,\tau_2) \triangleq \frac{f_2(\tau_2)-f_2(\tau_1)}{\tau_2-\tau_1}$ denotes the slope of the straight line passing through $(\tau_1,f_2(\tau_1))$ and $(\tau_2,f_2(\tau_2))$. Then, we show that all the points $(B_n,f_2(B_n))$ are above the straight line passing through such two points.

First, we find such two points $(\tau_1,f_2(\tau_1))$ and $(\tau_2,f_2(\tau_2))$ with $f_2'(\tau_1) = f_2'(\tau_2) = \theta(\tau_1,\tau_2)$ via the following procedure. To start with, we set $\hat\tau_1 = \zeta_1$ and $\hat\tau_2 = \zeta_2$. Since $f_2(B_n)$ is concave over $B_n \in[\zeta_1,\zeta_2]$, it is evident that $f'_2(\hat\tau_1) \ge \theta(\hat\tau_1,\hat\tau_2) \ge f_2'(\hat\tau_2)$, i.e., the slope of the line passing through $(\zeta_1,f_2(\zeta_1))$ and $(\zeta_2,f_2(\zeta_2))$ is between the values of $f_2'(\zeta_1)$ and $f_2'(\zeta_2)$. Then, we proceed as follows.
\begin{itemize}
  \item In the first step, we decrease the value of $\hat\tau_1$ to find a new $\hat\tau_1 > 0$ such that $f_2'(\hat\tau_1) = \theta(\hat\tau_1,\hat\tau_2)$. Note that $f_2(B_n)$ is convex over $B_n \in[0,\zeta_1]$, and thus decreasing $\hat\tau_1$ leads to the decrease of $f_2'(\hat\tau_1)$ and the increase of $\theta(\hat\tau_1,\hat\tau_2)$. Based on this fact together with $f_2'(0) = -\infty$, such a point $\hat\tau_1$ can be obtained via bisection. For this newly found $\hat\tau_1$, it follows that $f_2'(\hat\tau_1) = \theta(\hat\tau_1,\hat\tau_2) \ge f_2'(\hat\tau_2)$.
  \item In the second step, we increase the value of $\hat\tau_2$ to find a new $\hat\tau_2 > 0$ such that $\theta(\hat\tau_1,\hat\tau_2) = f_2'(\hat\tau_2)$. Note that $f_2(B_n)$ is convex over $B_n \in[\zeta_2,+\infty)$, and thus increasing $\hat\tau_2$ leads to the increase of $f_2'(\hat\tau_2)$ and the decrease of $\theta(\hat\tau_1,\hat\tau_2)$. Based on this fact together with $f_2'(+\infty) = 0$, such a point $\hat\tau_2$ can be obtained via bisection. For this newly found $\hat\tau_2$, it follows that $f_2'(\hat\tau_1) \ge \theta(\hat\tau_1,\hat\tau_2) = f_2'(\hat\tau_2)$.
  \item By iteratively implementing the above two steps, $f_2'(\zeta_1)$ is strictly increased and $f_2'(\zeta_2)$ is strictly decreased, while $\theta(\hat\tau_1,\hat\tau_2)$ is always between them. Note that $f_2(B_n)$ is continuous and second-order differentiable. By using this fact together with $f_2'(0) = -\infty$ and $f_2'(+\infty) = 0$, it is evident that there exist two finite extreme points $\tau_1$ and $\tau_2$, such that $f_2'(\tau_1) = \theta(\tau_1,\tau_2) = f_2'(\tau_2)$.
\end{itemize}

Next, we prove that all the points $(B_n,f_2(B_n))$ are above the line passing through $(\tau_1,f_2(\tau_1))$ and $(\tau_2,f_2(\tau_2))$. First, consider the regimes with $B_n \in[0, \zeta_1]$ and $B_n\in[\zeta_2,+\infty)$. Since the function $f_2(B_n)$ is convex over this regime, and $\theta(\tau_1,\tau_2) = f_2'(\tau_1) = f_2'(\tau_2)$, it is evident that over such two regimes, the points $(B_n,f_2(B_n))$ are above the line passing through $(\tau_1,f_2(\tau_1))$ and $(\tau_2,f_2(\tau_2))$. Then, consider the regime with $B_n\in [\zeta_1,\zeta_2]$. Since $f_2(B_n)$ is concave over this regime, the points $(B_n,f_2(B_n))$ are above the line passing through $(\zeta_1,f_2(\zeta_1))$ and $(\zeta_2,f_2(\zeta_2))$, and thus are also above that passing through $(\tau_1,f_2(\tau_1))$ and $(\tau_2,f_2(\tau_2))$.

By combining the above two steps, this lemma is proved.

\subsection{Proof of Proposition \ref{proposition:solution}}\label{proof:solution}

To start with, we define another auxiliary problem
\begin{align}
\min_{\{A_n \ge 0\},\{B_n\ge 0\}} & \frac{1}{2}\left(\mathbb{E}_{n\in\mathcal N_1} \left(f_1(A_n)\right) + \mathbb{E}_{n\in\mathcal N_2} \left(\bar f_2(B_n)\right)\right)\nonumber\\
\mathrm{s.t.}~~~~&~\frac{1}{2}\left(\mathbb{E}_{n\in\mathcal N_1} (A_n) + \mathbb{E}_{n\in\mathcal N_2} (B_n) \right)\le Q, \label{Problem:BPSK:4}
\end{align}
which is obtained based on problem (\ref{Problem:BPSK}) by replacing $f_2(B_n)$ as $\bar f_2(B_n)$. It is evident that the optimal value of problem (\ref{Problem:BPSK:4}) is a lower bound on that of problem (\ref{Problem:BPSK}). Therefore, if the objective value of problem (\ref{Problem:BPSK}) achieved by the solution in this proposition is same as the optimal value of problem (\ref{Problem:BPSK:4}), then such a solution is optimal for problem (\ref{Problem:BPSK}). We prove this proposition based on this observation.

First, we show that the optimal solution to problem (\ref{Problem:BPSK:4}) is given as $A_n^* = A^*, \forall n\in\mathcal N_1$ and $B_n^* = B^*, \forall n\in\mathcal N_2$. Note that both $f_1(A_n)$ and $f_2(B_n)$ are convex, and therefore, there exists an optimal power allocation solution in which the spoofing power $A_n$'s and $B_n$'s remain constant over $n\in\mathcal N_1$ and $n\in\mathcal N_2$, respectively. Therefore, we can express $A_n = A, \forall n\in\mathcal N_1$ and $B_n = B, \forall n\in\mathcal N_2$. Accordingly, problem (\ref{Problem:BPSK:4}) is degenerated to be problem (\ref{Problem:BPSK:5}). As a result, the optimal solution to problem (\ref{Problem:BPSK:4}) is $A_n^* = A^*, \forall n\in\mathcal N_1$ and $B_n^* = B^*, \forall n\in\mathcal N_2$.

Next, based on (\ref{eqn:barG}) and Lemma \ref{proposition:3}, it is easy to verify that the objective value of problem (\ref{Problem:BPSK}) achieved by the solution in this proposition is same as the optimal value of problem (\ref{Problem:BPSK:4}) achieved by $A_n^* = A^*, \forall n\in\mathcal N_1$ and $B_n^* = B^*, \forall n\in\mathcal N_2$. Therefore, this proposition is proved.

\subsection{Proof of Proposition \ref{proposition:QPSK:1}}\label{Appendix:E}

Consider one particular Type-I symbol with $x_n = \bar x_n = \frac{1 + j}{\sqrt{2}}$. Let the real and imaginary components of the spoofing signal $z_n$ be denoted as $z_n^{\rm R}$ and $z_n^{\rm I}$, and those of $y_n$ as $y_n^{\rm R}$ and $y_n^{\rm I}$, respectively. Then $y_n^{\rm R}$ and $y_n^{\rm I}$ are two real Gaussian random variables with mean values of $\sqrt{P/2} + z_n^{\rm R}$ and $\sqrt{P/2} + z_n^{\rm I}$, respectively, as well as variance of $1/2$. As a result, the joint PDF of $y_n^{\rm R}$ and $y_n^{\rm I}$ is given as
\begin{align}\label{eqn:joint:PDF}
p(y_n^{\rm R},y_n^{\rm I}) = \frac{1}{\sqrt{\pi}} e^{-(y_n^{\rm R}-\sqrt{P/2} - z_n^{\rm R})^2} e^{-(y_n^{\rm I}-\sqrt{P/2} - z_n^{\rm I})^2}.
\end{align}
Note that the spoofing is successful when the phase of $y_n$ lies between $0$ and $\pi/2$ (within the detection regime), i.e., the real and imaginary components of $y_n$ are both positive. Therefore, the conditional SSER under given $z_n^{\rm R}$ and $z_n^{\rm I}$ is given as
\begin{align}
&1- \int_{0}^{+\infty} \int_{0}^{+\infty} p(y_n^{\rm R},y_n^{\rm I})\mathrm{d} y_n^{\rm R} \mathrm{d} y_n^{\rm I} \nonumber\\
=& 1 - \int_{-\sqrt{P/2} - z_n^{\rm R}}^{+\infty} \int_{-\sqrt{P/2} - z_n^{\rm I}}^{+\infty} \frac{1}{\pi} e^{-{y_n^{\rm R}}^2} e^{-{y_n^{\rm I}}^2} \mathrm{d} y_n^{\rm R} \mathrm{d} y_n^{\rm I}\nonumber\\
=& 1- \left(\frac{1}{2}+\frac{1}{2}{\rm{erf}}\big(z_n^{\rm R}+\sqrt{P/2}\big)\right)\left(\frac{1}{2}+\frac{1}{2}{\rm{erf}}\big(z_n^{\rm I}+\sqrt{P/2}\big)\right).\label{eqn:g1}
\end{align}
To minimize the above conditional SSER under the given transmit power $A_n$, i.e., ${z_n^{\rm R}}^2+{z_n^{\rm I}}^2 = A_n$, it is desirable to set $z_n^{\rm R} \ge 0$ and $z_n^{\rm I}\ge 0$. As a result, obtaining $z_n^{\rm R}$ and $z_n^{\rm I}$ is equivalent to solving the following problem:
\begin{align}
\max_{z_n^{\rm R}\ge 0,z_n^{\rm I}\ge 0} &\ln \left(\frac{1}{2}+\frac{1}{2}{\rm{erf}}\big(z_n^{\rm R}+\sqrt{P/2}\big)\right) + \ln\left(\frac{1}{2}+\frac{1}{2}{\rm{erf}}\big(z_n^{\rm I}+\sqrt{P/2}\big)\right) \nonumber\\
\mathrm{s.t.}~~~&{z_n^{\rm R}}^2+{z_n^{\rm I}}^2 = A_n.\label{eqn:problem:tilde:f1}
\end{align}
Note that the function $\ln \left(\frac{1}{2}+\frac{1}{2}{\rm{erf}}\big(\sqrt{z}+\sqrt{P/2}\big)\right)$ is concave over $z \ge 0$. Therefore, by substituting $\hat z_n^{\rm R} = {z_n^{\rm R}}^2$ and $\hat z_n^{\rm I} = {z_n^{\rm I}}^2$ into problem (\ref{eqn:problem:tilde:f1}), we can show that the optimality is obtained as ${\hat z_n^{\rm R}} = { \hat z_n^{\rm I}} = {A_n/2}$. Therefore, the optimality of the problem (\ref{eqn:problem:tilde:f1}) is achieved when ${z_n^{\rm R}} = {z_n^{\rm I}} = \sqrt{A_n/2}$. By using this together with (\ref{eqn:g1}), the conditional SSER in (\ref{eqn:tilde:f1}) is obtained. Therefore, this proposition is proved.

\subsection{Proof of Proposition \ref{proposition:QPSK:2}}\label{Appendix:F}

Similar to the proof of Proposition \ref{proposition:QPSK:1}, the joint PDF of $y_n^{\rm R}$ and $y_n^{\rm I}$ is given in (\ref{eqn:joint:PDF}). Note that the spoofing is successful when the phase of $y_n$ lies between $\pi$ and $3\pi/2$, i.e., the real and imaginary components of $y_n$ are both negative. As a result, the conditional SSER ${\rm Pr}(\hat x_n\neq \bar x_n|x_n = \bar x_n =  (1+j)/\sqrt{2})$ is given as
\begin{align}
&1- \int_{-\infty}^{0} \int_{-\infty}^{0} p(y_n^{\rm R},y_n^{\rm I})\mathrm{d} y_n^{\rm R} \mathrm{d} y_n^{\rm I} \nonumber\\
=& 1- \left(\frac{1}{2}-\frac{1}{2}{\rm{erf}}\big(z_n^{\rm R}+\sqrt{P/2}\big)\right)\left(\frac{1}{2}-\frac{1}{2}{\rm{erf}}\big(z_n^{\rm I}+\sqrt{P/2}\big)\right).\label{eqn:g2}
\end{align}
To minimize the above conditional SSER under the given spoofing power $B_n$, i.e., ${z_n^{\rm R}}^2+{z_n^{\rm I}}^2 = B_n$, it is desirable to set $z_n^{\rm R} \le 0$ and $z_n^{\rm I}\le 0$. As a result, obtaining $z_n^{\rm R}$ and $z_n^{\rm I}$ is equivalent to solving the following problem:
\begin{align}
\max_{z_n^{\rm R} \le 0,z_n^{\rm I}\le0} &\ln \left(\frac{1}{2}-\frac{1}{2}{\rm{erf}}\big(z_n^{\rm R}+\sqrt{P/2}\big)\right) + \ln\left(\frac{1}{2}-\frac{1}{2}{\rm{erf}}\big(z_n^{\rm I}+\sqrt{P/2}\big)\right) \nonumber\\
\mathrm{s.t.}~&{z_n^{\rm R}}^2+{z_n^{\rm I}}^2 = B_n.\label{eqn:problem:tilde:f2}
\end{align}
By replacing $z_n^{\rm R}$ and $z_n^{\rm R}$ as $- \sqrt{\hat z_n^{\rm R}}$ and $- \sqrt{\hat z_n^{\rm I}}$, problem (\ref{eqn:problem:tilde:f2}) is recast as
\begin{align}
\max_{{\hat z_n^{\rm R}} \ge 0, {\hat z_n^{\rm R}}\ge0} &\ln \left(\frac{1}{2}+\frac{1}{2}{\rm{erf}}\big(\sqrt{\hat z_n^{\rm R}}-\sqrt{P/2}\big)\right) + \ln\left(\frac{1}{2}+\frac{1}{2}{\rm{erf}}\big(\sqrt{\hat z_n^{\rm I}}-\sqrt{P/2}\big)\right) \nonumber\\
\mathrm{s.t.}~&{\hat z_n^{\rm R}}+{\hat z_n^{\rm I}} = B_n.\label{eqn:problem:tilde:f2:2}
\end{align}
Note that when $P \le 4$, $\ln \left(\frac{1}{2}+\frac{1}{2}{\rm{erf}}\big(\sqrt{z}-\sqrt{P/2}\big)\right)$ is concave, and therefore, the optimality of problem (\ref{eqn:problem:tilde:f2:2}) is obtained as ${\hat z_n^{\rm R}} = { \hat z_n^{\rm I}} = {B_n/2}$. As a result, the optimality of problem (\ref{eqn:problem:tilde:f2}) is achieved when $z_n^{\rm R} =z_n^{\rm I} = -\sqrt{B_n/2}$. By using this together with (\ref{eqn:g2}), the conditional SSER in (\ref{eqn:tilde:f2}) is obtained. Therefore, this proposition is proved.

\subsection{Proof of Lemma \ref{lemma:QPSK:1}}\label{Appendix:G}

It is evident that $g_1(A_n)$ is monotonically decreasing over $A_n \ge 0$. Therefore, we only need to show its convexity. The second-order derivative of $g_1(A_n)$ is given as
\begin{align}
g''_1(A_n) = & A_n^{-1}e^{-\left(\sqrt{A_n/2}+\sqrt{P/2}\right)^2} \cdot
\left(-\frac{1}{4\pi}e^{-\left(\sqrt{A_n/2}+\sqrt{P/2}\right)^2} \right.\nonumber\\ &\left.+ \frac{1}{4\sqrt{2\pi}}(1+{\rm{erf}}(\sqrt{A_n/2} + \sqrt{P/2}))\left( A_n^{1/2} + \sqrt{P}+ A_n^{-1/2}\right)\right).
\end{align}
With $A_n \ge 0$, it follows that
\begin{align}
\frac{1}{4\sqrt{2\pi}}(1+{\rm{erf}}(\sqrt{A_n/2} + \sqrt{P/2}))\left( A_n^{1/2} + \sqrt{P}+ A_n^{-1/2}\right) & > \frac{1}{4\sqrt{2\pi}}A_n^{-1/2} >\frac{1}{4\pi}A_n^{-1/2}, \label{eqn:inequality:1}\\
\frac{1}{4\pi}e^{-\left(\sqrt{A_n/2}+\sqrt{P/2}\right)^2} &\le \frac{1}{4\pi}e^{-A_n/2}.\label{eqn:inequality:2}
\end{align}
Note that $A_n < e^{A_n}$ for all $A_n > 0$, and therefore, $\frac{1}{4\pi}e^{-A_n/2} < \frac{1}{4\pi}A_n^{-1/2}$. By using this together with (\ref{eqn:inequality:1}) and (\ref{eqn:inequality:2}), it follows that $
-\frac{1}{4\pi}e^{-\left(\sqrt{A_n/2}+\sqrt{P/2}\right)^2} + \frac{1}{4\sqrt{2\pi}}(1+{\rm{erf}}(\sqrt{A_n/2} + \sqrt{P/2}))\left( A_n^{1/2}+ \sqrt{P}+ A_n^{-1/2}\right) > 0$. Accordingly, $g_1''(A_n) > 0$ for all $A_n > 0$. As a result, $g_1(A_n)$ is a convex function, and this lemma is proved.

\subsection{Proof of Proposition \ref{lemma:QPSK:2}}\label{Appendix:H}

It is easy to see that $g_2(B_n)$ is monotonically decreasing over $B_n \in [0,+\infty)$. It thus remains to show its convexity. The second-order derivative of $g_2(B_n)$ is given as
\begin{align}
g''_2(B_n) &= B_n^{-1}e^{-\left(\sqrt{B_n/2}-\sqrt{P/2}\right)^2} \cdot \left(-\frac{1}{4\pi}e^{-\left(\sqrt{B_n/2}-\sqrt{P/2}\right)^2} \right.\nonumber\\
& \left. + \frac{1}{4\sqrt{2\pi}}(1+{\rm{erf}}(\sqrt{B_n/2}-\sqrt{P/2}))\left( B_n^{1/2} - \sqrt{P}+ B_n^{-1/2}\right)\right).
\end{align}

First, it is easy to see that $g''_2(B_n) \to +\infty$ as $B_n \to 0$, and $g''_2(\zeta_1) < 0$. Since $g_2(B_n)$ is a continuous function, there always exists a positive $\chi_1$ with $\chi_1 < \zeta_1$, such that over $B_n \in [0,\chi_1]$ we have $g''_2(B_n) \ge 0$, i.e., $g_2(B_n)$ is convex.

Next, note that when $B_n \ge P$, it follows that $1+{\rm{erf}}(\sqrt{B_n/2}-\sqrt{P/2}) \ge 1$. Also, when $B_n \ge \left(\frac{\sqrt{P} + \sqrt{\frac{\pi}{2}} + \sqrt{\left(\sqrt{P} + \sqrt{\frac{\pi}{2}}\right)^2-2}}{2}\right)^2$,  we have $\frac{1}{4\sqrt{2\pi}}\left( B_n^{1/2} - \sqrt{P}+ B_n^{-1/2}\right) \ge \frac{1}{4\pi}$. By combining the above two facts, when $B_n \ge \chi_2$, it holds that $ \frac{1}{4\sqrt{2\pi}}(1+{\rm{erf}}(\sqrt{B_n/2}-\sqrt{P/2}))\left( B_n^{1/2} - \sqrt{P}+ B_n^{-1/2}\right) \ge \frac{1}{4\pi} \ge \frac{1}{4\pi}e^{-\left(\sqrt{B_n/2}-\sqrt{P/2}\right)^2}$. Accordingly, $g''_2(B_n) \ge 0$ and $g_2(B_n)$ is convex.

Furthermore, when $P > 2$, it is easy to show that $B_n^{1/2} - \sqrt{P}+ B_n^{-1/2} \le 0$ for $B_n \in[\zeta_1,\zeta_2]$. Therefore, in this case $g''_2(B_n) \le 0$ and $g_2(B_n)$ is concave.


\begin{thebibliography}{1}

\bibitem{GCworkshop}
J. Xu, L. Duan, and R. Zhang, ``Transmit optimization for symbol-level spoofing with BPSK signaling,'' submitted to {\it IEEE GLOBECOM Workshop}.

\bibitem{ZouWangHanzo2015}
Y. Zou, X. Wang, and L. Hanzo, ``A survey on wireless security: technical challenges, recent advances and future trends,'' to appear in {\it Proc. IEEE}. [Online] Available: {\url{http://arxiv.org/abs/1505.07919}}.

\bibitem{XuDuanZhang1}
J. Xu, L. Duan, and R. Zhang, ``Proactive eavesdropping via jamming for rate maximization over Rayleigh fading channels,'' {\it IEEE Wireless Commun. Letters}, vol. 5, no. 1, pp. 80-83, Feb. 2016.

\bibitem{XuDuanZhang2}
J. Xu, L. Duan, and R. Zhang, ``Proactive eavesdropping via cognitive jamming in fading channels,'' in {\it Proc. IEEE ICC}, 2016.

\bibitem{ZengZhang}
Y. Zeng and R. Zhang, ``Active eavesdropping via spoofing relay attack,'' in {\it Proc. IEEE ICASSP}, 2016.

\bibitem{ZengZhang2}
Y. Zeng and R. Zhang, ``Wireless information surveillance via proactive eavesdropping with spoofing relay,'' to appear in {\it IEEE J. Sel. Topics Signal Process.}. [Online] Available: {\url{https://arxiv.org/abs/1606.03851}}.

\bibitem{Liu2015}
Q. Liu, M. Li, X. Kong, and N. Zhao, ``Disrupting MIMO communications with optimal jamming signal design,'' {\it IEEE Trans. Wireless Commun.}, vol. 14, no. 10, pp. 5313-5325, Oct. 2015.

\bibitem{Bayesteh2004}
A. Bayesteh, M. Ansari, and A. K. Khandani, ``Effect of jamming on the capacity of MIMO channels,'' in {\it Proc. IEEE 42nd Allerton Conf. Circuits and Systems Theory}, pp. 401-410, Oct. 2004.

\bibitem{Brady2006}
M. H. Brady, M. Mohseni, and J. M. Cioffi, ``Spatially-correlated jamming in Gaussian multiple access and broadcast channels,'' in {\it Proc. IEEE CISS}, pp. 1635-1639, Mar. 2006.

\bibitem{Rodrigues2009}
M. R. D. Rodrigues and G. Ramos, ``On multiple-input multiple-output Gaussian channels with arbitrary inputs subject to jamming,'' in {\it Proc. IEEE ISIT}, pp. 2512-2516, Jun. 2009.

\bibitem{Jorswieck2005}
E. A. Jorswieck, H. Boche, and M. Weckerle, ``Optimal transmitter and jamming strategies in Gaussian MIMO channels,'' in {\it Proc. IEEE VTC}, pp. 978-982, May 2005.

\bibitem{Medard}
M. Medard, ``Capacity of correlated jamming channels,'' in {\it Proc. 35th Allerton Conf.}, Monticello, IL, Oct. 1997, pp. 1043-1052.

\bibitem{Kashyap2004}
A. Kashyap, T. Basar, and R. Srikant, ``Correlated jamming on MIMO Gaussian fading channels,'' {\it IEEE Trans. Inf. Theory}, vol. 50, no. 9, pp. 2119-2123, Sep. 2004.


\bibitem{Shafiee2009}
S. Shafiee and S. Ulukus, ``Mutual information games in multi-user channels with correlated jamming,'' {\it IEEE Trans. Inf. Theory}, vol. 55, no. 10, pp. 4598-4607, Oct. 2009.


\bibitem{Nagarajan2010}
V. Nagarajan, V. Arasan, and D. Huang, ``Using power hopping to counter MAC spoof attacks in WLAN,'' in {\it Proc. IEEE CCNC}, Jan. 2010.

\bibitem{CERT1995}
Computer Emergency Response Team (CERT), ``CERT advisory: IP spoofing attacks and hijacked terminal connections.'' [Online] Available: {\url{http://www.cert.org/advisories/CA-1995-01.html}}, Jan. 1995.

\bibitem{Kannhavong2007}
B. Kannhavong, H. Nakayama, Y. Nemoto, N. Kato, and A. Jamalipour, ``A survey of routing attacks in mobile ad hoc networks,'' {\it IEEE Wireless Commun.}, vol. 14, no. 5, pp. 85-91, Dec. 2007.


\bibitem{Masouros2009}
C. Masouros and E. Alsusa, ``Dynamic linear precoding for the exploitation of known interference in MIMO broadcast systems,'' {\it IEEE Trans. Commun.}, vol. 8, no. 3, pp. 1396-1404, Mar. 2009.

\bibitem{Alodeh2015}
M. Alodeh, S. Chatzinotas, and B. Ottersten, ``Constructive multiuser interference in symbol level precoding for the MISO downlink channel,'' {\it IEEE Trans. Signal Process.}, vol. 63, no. 9, pp. 2239-2252, May 2015.


\bibitem{GlodsmithBook}
A. Goldsmith, {\it Wireless Communications.}  Cambridge, U.K.: Cambridge Univ. Press, 2004.

\end{thebibliography}
\end{document}